\documentclass[fleqn,usenatbib]{mnras}

\usepackage{newtxtext,newtxmath}

\usepackage[T1]{fontenc}

\DeclareRobustCommand{\VAN}[3]{#2}
\let\VANthebibliography\thebibliography
\def\thebibliography{\DeclareRobustCommand{\VAN}[3]{##3}\VANthebibliography}

\usepackage{graphicx}	
\usepackage{amsmath}	
\usepackage{multirow}

\title[High density relativistic reflection in RBS 1124 ]{Exploring the high-density reflection model for the soft excess in RBS\,1124}

\author[A. Madathil-Pottayil et al.]{A. Madathil-Pottayil,$^{1}$\thanks{E-mail: a.madathil-pottayil@herts.ac.uk}
D. J. Walton,$^{1}$
Javier García,$^{2,3}$
Jon Miller,$^{4}$
Luigi C. Gallo,$^{5}$
C. Ricci,$^{6,7}$
\newauthor
Mark T. Reynolds,$^{8,4}$
D. Stern,$^{9}$
T. Dauser,$^{10}$
Jiachen Jiang,$^{11,12}$
William Alston,$^{1}$
A. C. Fabian,$^{11}$
\newauthor
M. J. Hardcastle,$^{1}$
Peter Kosec,$^{13,14}$
Emanuele Nardini,$^{15}$
and Christopher S. Reynolds$^{16,17}$
\\
\\
$^{1}$Centre for Astrophysics Research, Department of Physics, Astronomy and Mathematics, University of Hertfordshire, College Lane, Hatfield AL10 9AB, UK\\
$^{2}$X-ray Astrophysics Laboratory, NASA Goddard Space Flight Center, Greenbelt, MD, USA\\
$^{3}$Cahill Center for Astronomy and Astrophysics, California Institute of Technology, Pasadena, CA 91125, USA\\
$^{4}$Department of Astronomy, University of Michigan, 1085 S. University Ave., Ann Arbor, MI 48109, USA\\
$^{5}$Department of Astronomy and Physics, Saint Mary’s University, 923 Robie Street, Halifax, NS B3H 3C3, Canada\\
$^{6}$Instituto de Estudios Astrofísicos, Facultad de Ingeniería y Ciencias, Universidad Diego Portales, Av. Ejército Libertador 441, Santiago, Chile \\
$^{7}$Kavli Institute for Astronomy and Astrophysics, Peking University, Beijing 100871, People's Republic of China\\
$^{8}$Department of Astronomy, Ohio State University, 140 W. 18th Avenue, Columbus, OH 43210, USA\\
$^{9}$Jet Propulsion Laboratory, California Institute of Technology, 4800 Oak Grove Drive, Mail Stop 264-789, Pasadena, CA 91109, USA\\
$^{10}$Dr. Karl-Remeis-Sternwarte and ECAP, Friedrich-Alexander-Universität Erlangen-Nürnberg, Sternwartstr. 7, 96049 Bamberg, Germany\\
$^{11}$Institute of Astronomy, University of Cambridge, Madingley Road, Cambridge CB3 0HA, UK\\
$^{12}$Department of Physics, University of Warwick, Gibbet Hill Road, Coventry CV4 7AL, UK\\
$^{13}$Center for Astrophysics, Harvard \& Smithsonian, 60 Garden St, Cambridge, MA 02138, USA\\
$^{14}$Kavli Institute for Astrophysics and Space Research, Massachusetts Institute of Technology, MA, USA\\
$^{15}$INAF – Osservatorio Astrofisico di Arcetri, Largo Enrico Fermi 5, 50125 Firenze, Italy\\
$^{16}$Department of Astronomy, University of Maryland, College Park, MD20742, USA\\
$^{17}$Joint Space Science Institute (JSI), University of Maryland, College Park, MD20742, USA
}
\date{Accepted 2024 August 30. Received 2024 August 7; in original form 2024 March 25}

\pubyear{2024}

\begin{document}
\label{firstpage}
\pagerange{\pageref{firstpage}--\pageref{lastpage}}
\maketitle

\begin{abstract}
$`$Bare' active galactic nuclei (AGN) are a subclass of Type 1 AGN that show little or no intrinsic absorption. They offer an unobscured view of the central regions of the AGN and therefore serve as ideal targets to study the relativistic reflection features originating from the innermost regions of the accretion disc. We present a detailed broadband spectral analysis (0.3 -- 70 keV) of one of the most luminous bare AGN in the local universe, RBS\,1124 ($z= 0.208$) using a new, co-ordinated high signal-to-noise observation obtained by \textit{XMM-Newton} and \textit{NuSTAR}. The source exhibits a power-law continuum with $\Gamma \sim$ 1.8 along with a soft excess below 2 keV, a weak neutral iron line and curvature at high energies ($\sim 30$ keV). The broadband spectrum, including the soft excess and the high-energy continuum, is well fit by the relativistic reflection model when the accretion disc is allowed to have densities of log$(n_{\rm e}$/cm$^{-3}$) $\gtrsim 19$. Our analysis therefore suggests that when high-density effects are considered, relativistic reflection remains a viable explanation for the soft excess.
\end{abstract}

\begin{keywords}
black hole physics – galaxies: active – X-rays: individual: RBS\,1124
\end{keywords}



\section{Introduction}
\label{sec-1}

The X-ray spectra of Type 1 AGN are known to exhibit a \textit{soft\, excess} below 2 keV, a feature characterized by a smooth excess flux over the primary continuum when extrapolated down to lower energies \citep{1985MNRAS.217..105A,1995MNRAS.277L...5P,1999ApJS..125..297L}. The origin of soft excess is still not clear and has been one of the long-standing mysteries in X-ray studies of AGN. It was initially believed to be the high-energy tail of blackblody emission from the inner disc \citep{1985MNRAS.217..105A,1998MNRAS.301..179M}. However, the disc temperatures derived from the models (0.1 -- 0.3 keV) were deemed too hot to be considered as thermal emission from the accretion disc of an AGN \citep{2004A&A...422...85P,2009MNRAS.394..443M}. In addition, the soft excess `temperature' remains constant across several orders of magnitude of black hole mass and accretion rate (\citealt{2006MNRAS.365.1067C,2006MNRAS.368..903P,2012MNRAS.420.1848D}). These findings essentially rule out thermal emission from the disc as the primary origin. Alternatively, it was later proposed that the soft excess could be produced by a smeared ionized relativistic absorption model, where the absorption features (oxygen edges) due to ionized relativistic winds originating in the  inner parts of the accretion disc are broadened due to the relativistic motions of the wind (\citealt{2004MNRAS.349L...7G,2007MNRAS.381.1426M}). This generates a deficit at $\sim$ 1 keV, creating the appearance of an excess at soft energies and spectral hardening at higher energies. However, simulations of such a wind model in \cite{2007MNRAS.381.1413S} show sharp features in the spectra and their calculations reveal that the winds do not achieve high enough relativistic speeds to produce absorption that is sufficiently smeared to explain the soft excess.

Currently there are two prevailing theories that are commonly used to describe the soft excess: the warm corona model \citep{1998MNRAS.301..179M,2012MNRAS.420.1848D,2013A&A...549A..73P} and the blurred relativistic reflection model \citep{2006MNRAS.365.1067C,2008MNRAS.391.2003Z,2013MNRAS.428.2901W}.  Each of these models adopt different physical mechanisms to explain the soft excess. The warm corona model postulates the soft excess to be the result of Compton upscattering of disc photons from an additional corona, which is much cooler ($kT_{e} \sim 0.1$ keV) and has a much higher optical depth ($\tau = 20 - 40$) compared to the `hot' corona ($kT_{e} > 50$ keV, $\tau \lesssim 1$) that generates the primary continuum. The blurred reflection model, on the other hand, implements the effects of special and general relativity \citep{2000PASP..112.1145F} on the forest of emission lines within the reflection spectrum from an ionized disc at energies $< 2$ keV \citep{2005MNRAS.358..211R,2013ApJ...768..146G}, originating close to the innermost stable circular orbit of the supermassive black hole (SMBH). The blurred emission lines blend together, thereby reproducing the soft excess. The reflection origin scenario for the soft excess is further supported by the detection of soft lags, where the soft photons lag behind the hard X-ray continuum, as a result of the delay due to reverberation from the accretion disc \citep{2009Natur.459..540F,2012MNRAS.422..129Z,2016MNRAS.462..511K}. 

At present, it remains unclear which model is the primary origin of the soft excess. \cite{2019ApJ...871...88G} argued that the relatively cool temperature of the warm corona should result in numerous absorption features in the soft X-ray band, which do not seem to be observed. However, \cite{2020A&A...634A..85P} later argued that the warm corona could remain completely ionized, thereby exhibiting no absorption/emission features in the soft energy band, if heating from the hot corona, the disc below, and internal heating processes are considered in their radiative transfer code, TITAN. While the debate between the warm corona and relativistic reflection models persists, some studies suggest that a combination of both models provides a more accurate explanation of the soft excess feature \citep{2018A&A...609A..42P,2021A&A...654A..89P,2021ApJ...913...13X,2022MNRAS.515..353X}.

In the context of a reflection scenario for the soft excess, it was recently shown that reflection from discs with densities higher than $10^{15}$ cm$^{-3}$ can contribute towards the soft excess in a second manner as well. Reflection models by \cite{2016MNRAS.462..751G} demonstrated that in a disc with higher density ($n_{\rm e} > 10^{15}$ cm$^{-3}$) there is a higher probability for free-free absorption, resulting in a rise in the disc temperature. Moreover, due to the quadratic dependence of free-free heating on density, the free-free continuum emissivity is enhanced, leading to a consequent increase in the reflection continuum. With this increase in the reflection continuum, along with the temperature rise in the disc, gas on the surface of the disc radiates like a black body, which can make a further contribution to the soft excess in the reflection interpretation (in addition to the blurred emission lines). Integrating variable disc densities with blurred relativistic reflection creates the high disc density relativistic reflection model, a promising tool for observing and understanding disc density. This high-density reflection model has successfully explained the soft excess component in several AGN without the requirement of any additional Comptonizing component \citep{2019MNRAS.489.3436J,2021ApJ...913...13X,2022MNRAS.513.4361M,2022MNRAS.514.1107J,2023MNRAS.522.5456Y}. 

However, most of these high-density reflection studies have focused on the soft X-ray band ($0.3 - 10$ keV). There have been only a few instances where the model was tested with broadband datasets of `bare' AGN 
(which have little-to-no excess absorption above the Galactic column, and thus provide the best view of the soft excess; \citealt{2020MNRAS.497.2352M,2021A&A...654A..89P,2021ApJ...913...13X,2022MNRAS.517.4788C}). This limited testing has resulted in a significant ongoing debate as to whether the high density relativistic reflection model can adequately fit the broadband data for these sources. Some analyses have suggested that, while the model can successfully fit data in the soft band, in some cases it appears to struggle to fit the full broadband spectrum. For example, \cite{2020MNRAS.497.2352M} reported that the high-density reflection model, though able to fit the soft X-ray spectrum of Ton S180 with moderate density values \citep{2019MNRAS.489.3436J}, struggled to simultaneously explain both the soft excess and features beyond 10 keV when \textit{NuSTAR} data was included. \cite{2021A&A...654A..89P} also report similar results when applying the model to Mrk 110. In contrast, for some bare AGN the relativistic reflection model can successfully model the available broadband data (e.g. \citealt{2019ApJ...871...88G,2021ApJ...913...13X}). Given these circumstances, it is crucial to continue testing the high-density reflection model on broadband observations to fully understand its potential. In this context, we decided to assess the high disc density relativistic reflection model using the bare AGN RBS 1124.

RBS 1124 ($z = 0.208$), also known as RX J1231.6+7044, was originally detected as an X-ray source during the \textit{ROSAT All Sky Survey (RASS)}, and is one of the most luminous radio-quiet quasars in the local Universe ($L_{\mathrm 2-10~\rm keV} = 6\times10^{44}$ ergs s$^{-1}$; \citealt{2010MNRAS.401.1315M}). It is categorized as a bare AGN with no/minimal intrinsic obscuration (\citealt{2013MNRAS.428.2901W}), thereby offering an unobscured view of the central regions of the AGN and an ideal opportunity to study relativistic reflection features originating from the innermost regions of the accretion disc. RBS 1124 is spectroscopically classified as a broad line QSO with an H$\beta$ line width of $(4.26\pm{1.25}) \times10^{3}$ km s$^{-1}$ (full width at half maximum; \citealt{1998A&A...335..467T}) and an optical luminosity at $5100$ \AA  ~of $L_{5100} = 1.2 \times 10^{44}$ erg s$^{-1}$. \cite{2010MNRAS.401.1315M} estimate the mass of the black hole to be around $1.8 \times 10^{8}M_\odot$ using the mass scaling relationship between H$\beta$ line width and $L_{5100}$ continuum luminosity (\citealt{2006ApJ...641..689V}).

Previous analysis of the \textit{Suzaku} observation of RBS 1124 showed that the source had a moderately hard power law continuum with photon index ($\Gamma$) of $1.94\pm0.02$, a soft excess, a broad Fe-K$\alpha$ fluorescence line in the $0.5 - 10$ keV band (\citealt{2010MNRAS.401.1315M,2010ApJ...725.2444K}), a Compton hump peaking between $15 - 25$ keV in the hard X-ray band (\citealt{2013MNRAS.428.2901W, 2020MNRAS.498.5207W}), and  radiatively efficient accretion with $L_{\rm Bol}/L_{\rm Edd} \simeq 0.15$. Although \textit{Suzaku} offered a broadband view of the source, \textit{Suzaku}/PIN is a non-imaging high-energy detector (10 -- 70 keV) which is dominated by background, hence providing a low signal-to-noise (S/N) ratio. 
Observations with high S/N broadband coverage, combining \textit{NuSTAR} and \textit{XMM-Newton} for example, are vital for determining the presence of relativistic reflection from the disc, and for placing the most robust constraints on its properties (\citealt{2013Natur.494..449R,2014ApJ...788...61B,2014ApJ...788...76W,2015MNRAS.449..234K,2016MNRAS.456L..94M,2018MNRAS.480.3689B}).  In this paper, we present a detailed broadband spectral analysis (0.3 -- 70 keV) of RBS 1124 using a new, co-ordinated high S/N observation obtained with \textit{XMM-Newton} and \textit{NuSTAR}. 

In the following sections we introduce the observations used for our analysis (Section \ref{sec-2}). The data analysis procedures and the modelling techniques are described in Section \ref{sec-3}. In Section \ref{sec-4} and Section \ref{sec-5} we present the results from our analysis and discuss their implications, respectively.

\section{Observations and Data Reduction}
\label{sec-2}

\begin{figure}
\centering
\hspace*{-0.2cm}\includegraphics[scale=0.352]{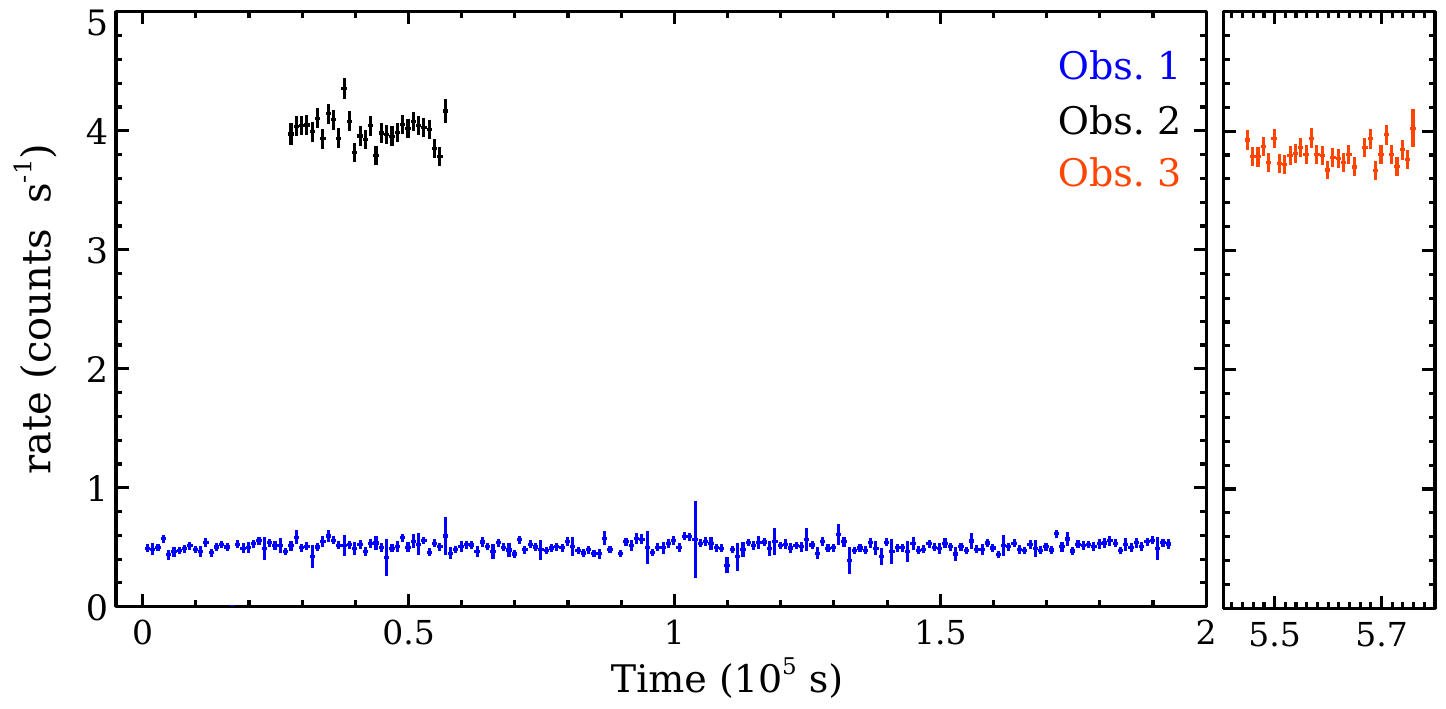}
\caption{\textit{NuSTAR} (FPMA+FPMB) full band light curve for Obs. 1 with an exposure of $135$ ks is plotted in blue. \textit{XMM-Newton} (EPIC-PN) light curves for Obs. 2 and 3, in the full energy band, each having an exposure of $\sim$ 30 ks, are plotted in black and orange colors, respectively. All the light curves are binned to 1000 s}
\label{fig-1}
\end{figure}

We use three new, co-ordinated high S/N observations obtained from \textit{XMM-Newton} and \textit{NuSTAR} taken during 2021. The log of the observations are given in Table \ref{tab-1}. During this period the source did not show any specific variability (see Figure \ref{fig-1}) and so we focus our analysis on modelling the time-averaged spectrum obtained from these observations. We carried out a standard data reduction of the three observations and we analyse the data using two of the leading relativistic reflection models, which will be described in Sections \ref{sec-3.1} and \ref{sec-3.2}. 

\begin{table*}
\centering
\caption{Summary of the observations of RBS 1124}
\begin{tabular}{cccccc}
\hline
Obs. & Mission & Instrument &  Observation & Start Date & Exposure time (s) \\
\hline
 1 & \textit{NuSTAR} & FPMA \& FPMB & 60701055002 & 14/12/2021 & 135952\\
 2 & \textit{XMM-Newton} & PN, MOS \& RGS & 0891804001 & 14/12/2021 &  30312\\
 3 & \textit{XMM-Newton} & PN, MOS \& RGS & 0891804201 & 20/12/2021 & 32117\\
\hline
\end{tabular}
\label{tab-1}
\end{table*}

\subsection{XMM-Newton}
\label{sec-2.1}
We used observations from the European Photon Imaging Camera (EPIC) and the Reflection Grating Spectrometer (RGS) instruments onboard \textit{XMM-Newton}. EPIC includes three X-ray CCD cameras covering the 0.3 -- 10 keV bandpass: two Metal Oxide Semi-conductors (MOS1 and MOS2; \citealt{2001A&A...365L..27T}) and a p-n CCD (PN; \citealt{2001A&A...365L..18S}). The two RGS detectors (RGS 1 \& 2) provide dispersive high-resolution spectroscopy over the 0.35 -- 2 keV energy range \citep{2001A&A...365L...7D}.

The Observation Data Format (ODF) files for both the \textit{XMM-Newton} observations were retrieved from the \textit{XMM-Newton} Science Archive (XSA) and reduced using the standard software package, the Science Analysis System (SAS), \textsc{sasversion} 20.0.0, maintained by the Science Operations Centre (SOC). All EPIC detectors were operated in Small Window mode for both observations. We then ran the reduction meta-tasks \textsc{emproc} and \textsc{epproc} to obtain the calibrated and integrated event lists for the MOS and PN detectors, respectively. The \textsc{xmmselect} task was  invoked to generate the science products from these cleaned event lists. For the PN observations, a source region of 30$^{\prime\prime}$ and a background region of 60$^{\prime\prime}$ were chosen. For MOS observations, a source region of 40$^{\prime\prime}$ and a background region of 160$^{\prime\prime}$ were chosen. The standard additional filters were used during event selection: we only select events with PATTERN $\leq 4$ (single-double events) and PATTERN $\leq 12$ (single-quadruple events) for the PN and MOS detectors, respectively, and ran our extraction with FLAG $==0$ to exclude events close to CCD gaps and/or bad pixels. We used the \textsc{epatplot} task to check whether there is any pile-up in the data, and this was found to be negligible. The response files for both PN and MOS detectors were generated using the \textsc{rmfgen} and \textsc{arfgen} tasks. We thus obtained three sets of spectral files (from the PN, MOS1 and MOS2 detectors) for each observation. We also checked the spectra to see if the source exhibited any significant variability between the two observations. The overall shape of the spectra were similar without any significant variability. We therefore combined the PN spectra and the spectra from the MOS detectors (MOS1 and 2) for both epochs using the \textsc{addascaspec} task, resulting in a single set of spectral files for the PN detector, and a single set of spectral files for the combined MOS detectors. The final PN and MOS spectra were rebinned to have a minimum S/N of 10 per energy bin.

The RGS data were reduced using standard procedures, with both spectrometers operating in the `Spectroscopy’ mode for both observations. Clean event lists, light curves, spectra, and response files were generated using the \textsc{rgsproc} task.  Standard source and background extraction regions were used. The spectra from RGS 1 and 2 of both Obs 2 and 3 were merged into a single spectrum using the \textsc{rgscombine} task, after ensuring consistency in overall spectral shape across all four RGS datasets. To maintain maximum spectral resolution, the RGS data was binned to a minimum of 1 count per bin.

\subsection{NuSTAR}
\label{sec-2.2}
\textit{NuSTAR} \citep{2013ApJ...770..103H} carries two identical, co-aligned X-ray telescopes that are sensitive over the 3--78\,keV band (focal plane modules A and B; FPMA and FPMB). The data for both FPMA and FPMB were reduced with the \textit{NuSTAR} Data Analysis
Software \textsc{(nustardas) v}2.0.0, and \textit{NuSTAR} calibration database v20221115. The unfiltered event files were initially cleaned with \textsc{nupipeline}. We used the standard depth correction to reduce the
internal high-energy background, and passages through the South Atlantic Anomaly were removed using the following settings: \textsc{saacalc = 3}, \textsc{saa = optimized} and \textsc{tentacle = yes}. Source lightcurves, spectra and their associated instrumental responses were then extracted from these cleaned event lists using \textsc{nuproducts} and a circular aperture of radius 70$''$. The background contribution was estimated from a larger region of blank sky on the same chip as RBS 1124. In order to maximise the S/N, we extracted both the standard `science' data (mode 1) and the `spacecraft science' data (mode 6), following the procedure outlined by \cite{2016ApJ...826...87W} in the latter case; the mode 6 data provide 19\% of the total good exposure for this observation. After performing the spectral extraction for FPMA and FPMB separately, these data were combined into a single \textit{NuSTAR} spectrum using \textsc{addascaspec}, and the
data were rebinned to have a minimum S/N of 5 per energy bin. 
The source is detected over the 3–70 keV energy range. Considering its redshift, it is worth noting that this \textit{NuSTAR} detection extends to an intrinsic spectral energy of up to 85 keV, indicating a significant detection.

\section{Spectral Analysis}
\label{sec-3}

In this section, we analyse the time-averaged spectrum in the broadband energy range (0.3 -- 70 keV) and describe the  model-setup procedure for fitting the spectrum. \textsc{xspec v}12.12.1 \citep{1996ASPC..101...17A} is used to analyse  our data and the errors we quote are the 90\% confidence interval for a single parameter of interest. Throughout this work, we use the \textit{tbabs} neutral absorption code \citep{2000ApJ...542..914W} to model neutral hydrogen absorption within our Galaxy along the line of sight. The equivalent neutral hydrogen column density value ($N_{\mathrm H}$) is fixed to 1.5$\times10^{20}$ cm$^{-2}$ based on \cite{2016A&A...585A..41W}. We also include a \textit{constant} component to account for the cross-normalization  factors between the three detectors ({PN, MOS and \textit{NuSTAR}}) for the joint fit.

\begin{figure*}
  \centering
  \begin{minipage}[b]{0.35\textwidth}
       \hspace*{-0.15cm} \includegraphics[scale=0.35]{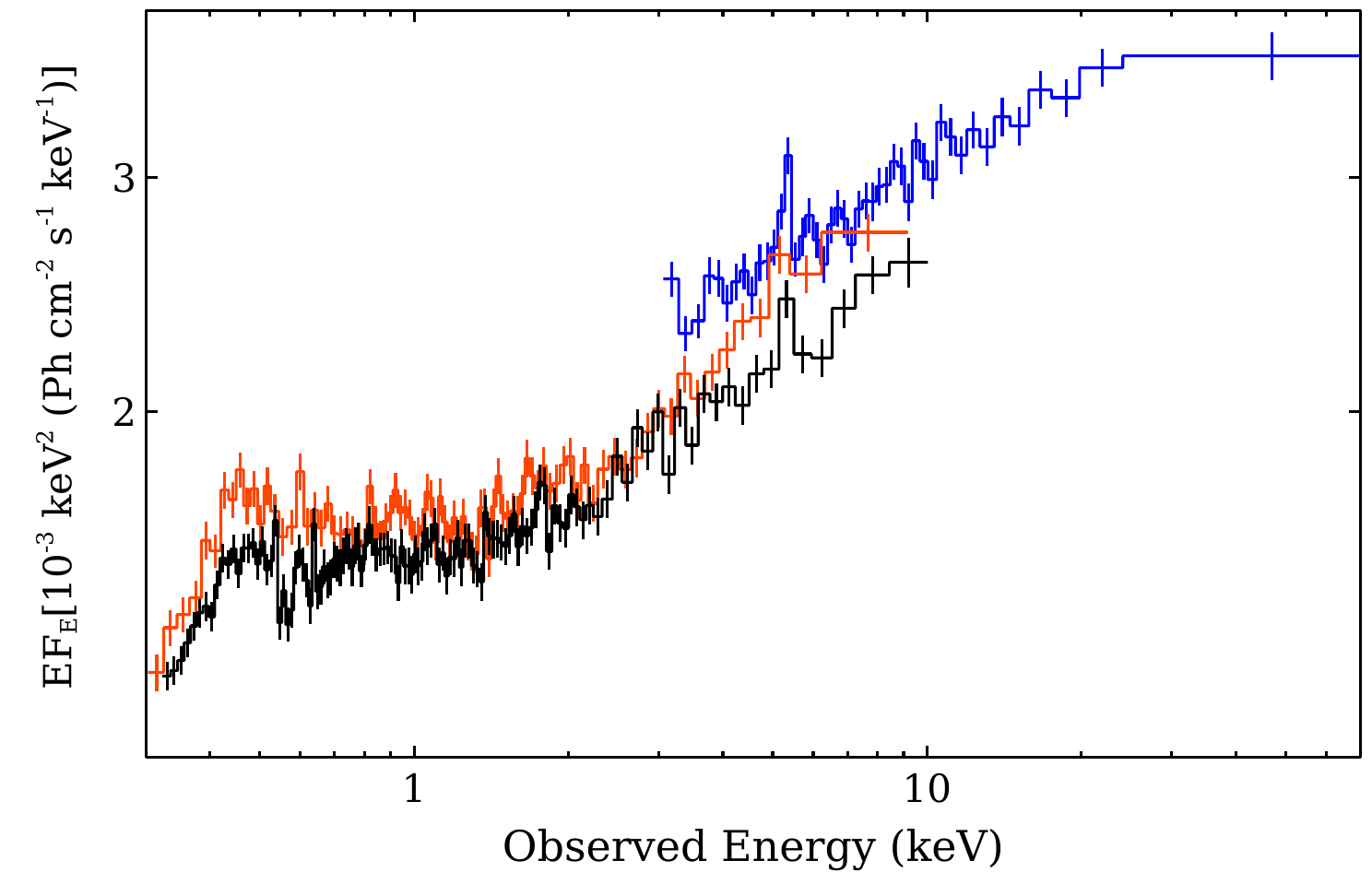}
     \label{fig2a}
  \end{minipage}
  \hfill
    \begin{minipage}[b]{0.35\textwidth}
  \hspace*{-2.55cm}
    \includegraphics[scale=0.35]{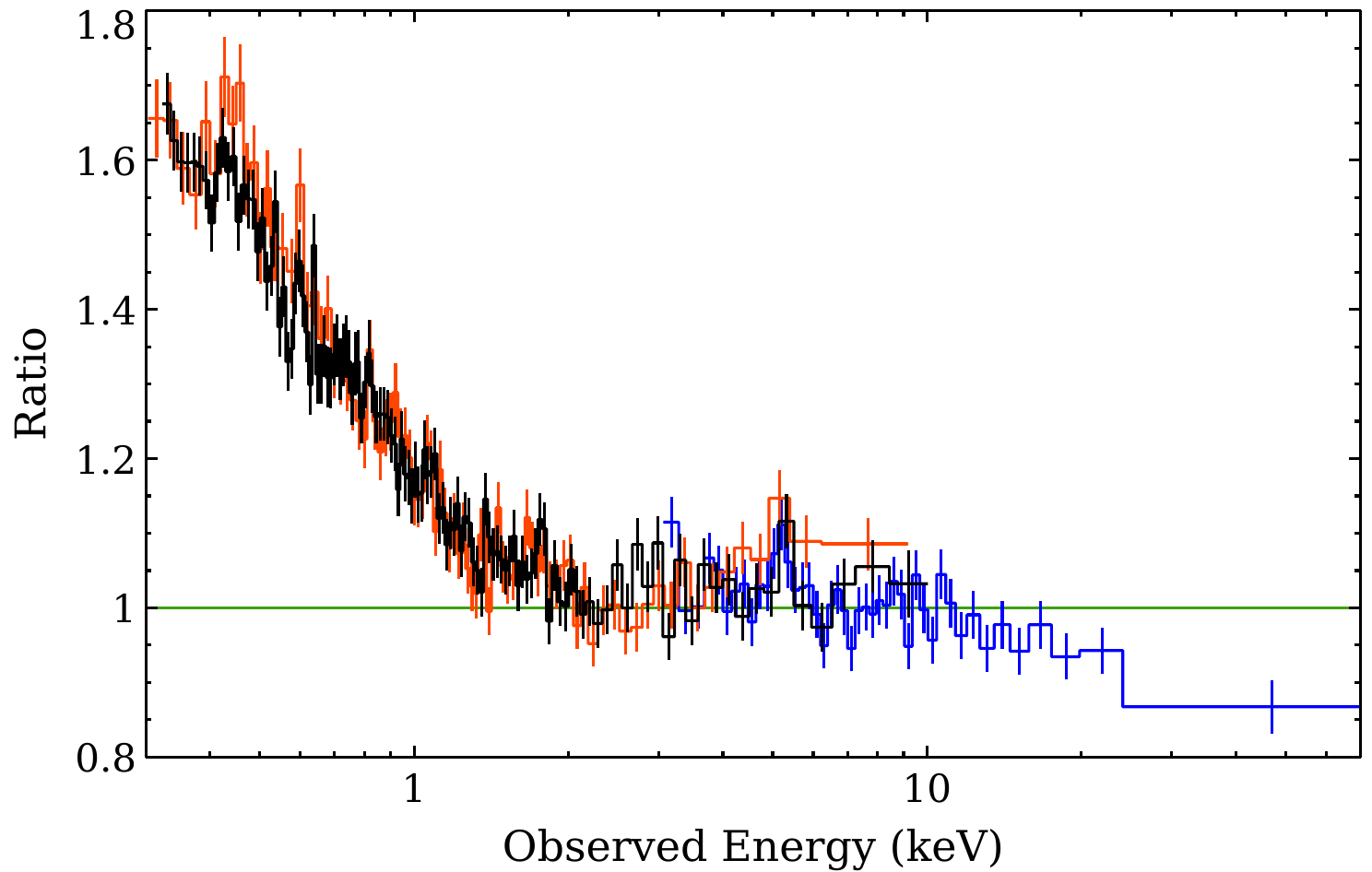}
    \label{fig2b}
  \end{minipage}
  
  \caption{\textit{Left}: The broadband spectrum of RBS 1124 unfolded through a simple powerlaw model. The data plotted in black, orange and blue  shows the PN, MOS and \textit{NuSTAR} data, respectively.  \textit{Right}: Residuals obtained when plotting the data with respect to an absorbed powerlaw. The data is fitted over the energy ranges 2 -- 4 keV, 7 -- 10 keV and 30 -- 70 keV and then extrapolated for the broadband energies. The key signatures include the weak neutral iron line, soft excess and curvature at higher energies associated with a high energy cutoff. \label{fig-2}}
\end{figure*}

 The unfolded \textit{XMM-Newton$+$NuSTAR} spectrum of RBS 1124, observed during December 2021, is shown in the left panel of Figure \ref{fig-2}. To start, we employed a simple \textit{powerlaw} model modified by Galactic absorption which we fit to the energy ranges of 2 -- 4 keV, 7 -- 10 keV, and 30 -- 70 keV to emphasize the spectral features above the power-law continuuum; these are the energy ranges in which the primary AGN continuum is expected to be the dominant emission component. We then extrapolate this fit to the full bandpass considered here (0.3 -- 70 keV). The resulting data/model ratio is shown in the right panel of Figure \ref{fig-2}. The source spectrum is characterized by a primary continuum of $\Gamma = 1.72$. The source shows a soft excess below 1.5 keV with reference to the \textit{powerlaw}. We also detect a faint excess emission at $\sim 5.4$ keV (observed frame), which is associated with the neutral iron line after accounting for the cosmological redshift of RBS 1124. Along with these features, we also observe curvature at high energies, which may indicate that we see the high-energy cutoff related to the electron temperature of the corona. Based on the residuals, we next investigate the presence of iron emission  and update the broadband X-ray spectral model.

\subsection{RGS spectral analysis }

Before proceeding with the broadband analysis of the source, we first present a brief analysis of the RGS data to confirm the `bare’ nature of RBS 1124. The soft excess component is dominant across the energy range covered by RGS ($0.35-2$ keV), as seen in the right panel of Figure \ref{fig-2}. The merged RGS spectrum was initially modelled using the phenomenological baseline model, combining \textit{powerlaw} and \textit{blackbody} components to address the primary continuum and the soft excess, respectively. However, the \textit{blackbody} contribution was unconstrained, and the continuum was well-described by the \textit{powerlaw} alone, with no improvement from the \textit{blackbody} inclusion. We therefore use the simpler \textit{powerlaw} continuum modified by Galactic absorption.

\begin{figure}
\centering
\hspace*{-0.4cm}
\includegraphics[scale=0.35]{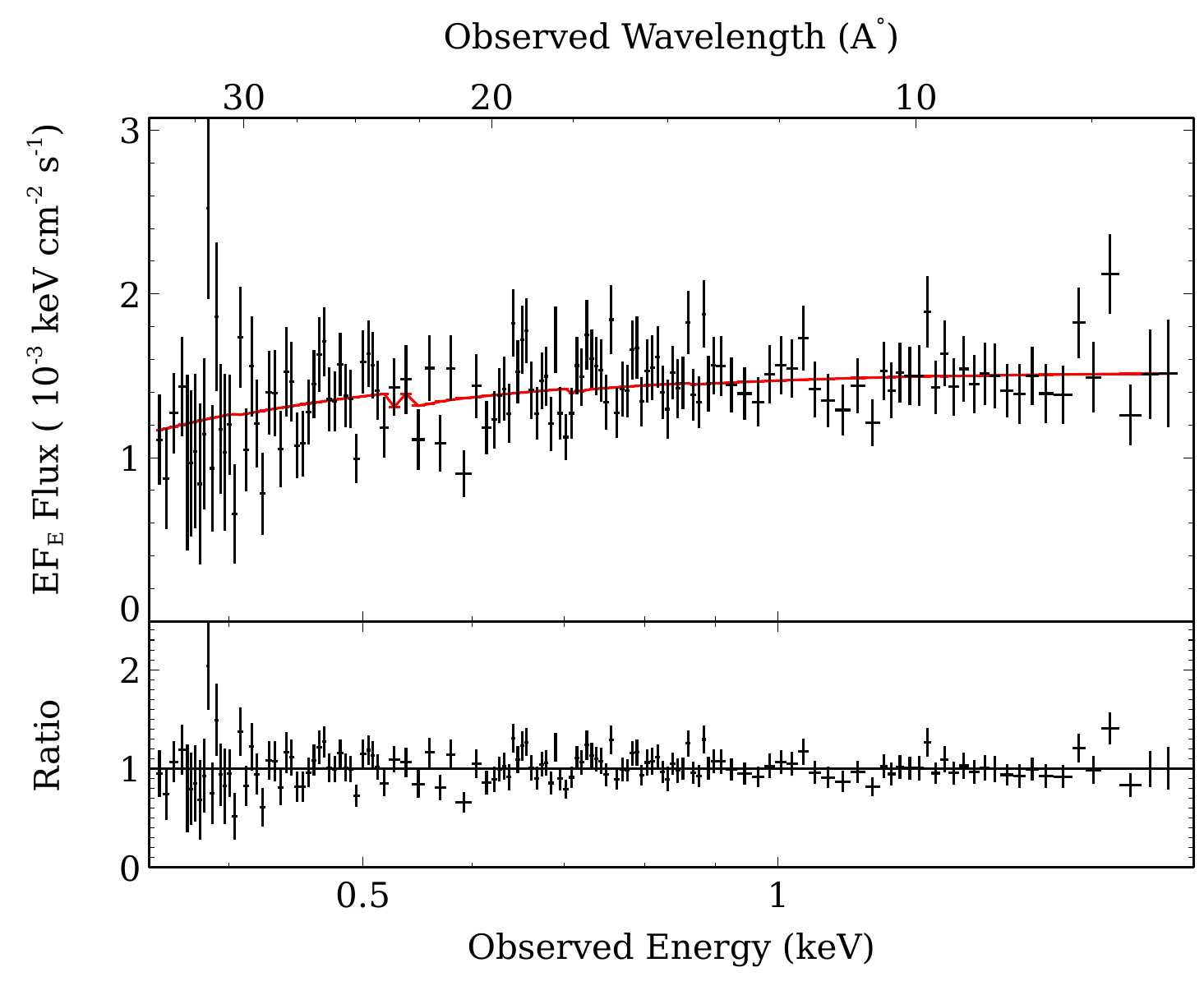}
\caption{ The spectral fit of RGS data using a simple \textit{powerlaw} modified by Galactic absorption is shown in the upper panel. The black points represent the data, and the red solid line represents the model. The bottom panel illustrates the residuals against the absorbed \textit{powerlaw} model. The RGS spectrum has been rebinned for plotting purposes.}
\label{fig-3}
\end{figure}

Due to the low count rates per channel in the RGS data, we use the Cash statistic \citep{1979ApJ...228..939C} for spectral fitting. This model produced a good fit, with a C-statistic of 2834.5 for 2813 degrees of freedom (d.o.f) and a $\Gamma$ of 2.00$\pm$0.05\footnote{The \textit{powerlaw} now accounts for the soft excess over the RGS bandpass, resulting in a steeper $\Gamma$ value than the one obtained in the broadband analysis later on (see Section \ref{sec-4}). For completeness, we also tried fixing $\Gamma$ to the value implied by the broadband analysis ($\Gamma = 1.8$) and reintroducing the \textit{blackbody} component for the soft excess. This did not change any of the conclusions regarding the presence of the absorption obtained with the simpler \textit{powerlaw} continuum.}. There were no obvious wind signatures in the RGS spectrum. However, slight residuals at $\sim0.6$ keV were observed (as shown in Figure \ref{fig-3}). To investiagte these residuals, we added the multiplicative \textit{XSTAR} table model to the \textit{powerlaw} model. This addition resulted in a marginal improvement with a $\Delta$C-statistic of -12 for 3 d.o.f.

The best-fit model parameters correspond to a moderately-ionized, low column density, ultrafast outflowing wind, with log ($\xi$ erg cm s$^{-1}$) $= 1.6\pm0.2$, $N_{\rm H} =  3.7^{+3.0}_{-2.4}\times10^{20}$ cm$^{-2}$ and $\nu = 0.09\pm0.01c$, respectively. However, this marginal improvement in the fit statistics is not significant. Moreover, the combination of such a fast outflow velocity with a low ionization parameter is unusual, and not typical of most of the `ultrafast’ outflows claimed in literature \citep{2010A&A...521A..57T,2013MNRAS.430...60G,2018ApJ...854L...8R}. Therefore, we also performed fits with the outflow velocity set to zero and the ionization parameter capped at an upper limit of 3 to closely mimic a typical warm absorber \citep{2007MNRAS.379.1359M,2014MNRAS.441.2613L,2023arXiv230210930G}. We found that including an ionized absorber with low outflow velocity did not improve the fit at all. The upper limit on the column density was  constrained to $N_{\rm H}< 6\times10^{21}$ cm$^{-2}$, placing RBS 1124 at the lower end of the $N_{\rm H}$ distribution seen in warm absorbers from other systems \citep{2014MNRAS.441.2613L,2024arXiv240502391Y}, thereby affirming the `bare’ nature of RBS\,1124.

No significant emission lines were detected in the RGS spectrum, including the O~\textsc{vii}  triplet recently reported in similar bare AGN \citep{2016ApJ...828...98R,2024arXiv240613623P}. Assuming a line width ($\sigma$) of 3.5 eV\footnote{The line width is fixed at $\sigma = 3.5$ eV based on the broad lines observed in this source (FWHM (H$\beta$) $\sim 4250$ $\rm km~s^{-1}$; \citealt{2004AJ....127..156G}), which is also consistent with the widths of the lines from the O~\textsc{vii} triplet in \cite{2016ApJ...828...98R} and \cite{2024arXiv240613623P}.}, the most conservative upper limit on the equivalent width (EW) of the O~\textsc{vii} triplet emission was constrained to EW $< 3.8$ eV for a forbidden transition at a rest-frame energy of 561 eV. This limit is comparable to the constraints reported by \cite{2016ApJ...828...98R} and \cite{2024arXiv240613623P} for the O~\textsc{vii} intercombination and resonant lines, respectively, with EWs $<3.3$ eV and $<3.1$ eV. This suggests that while features of similar strength could be consistent with our data, we do not detect them in RBS 1124.

\subsection{Iron Emission}
\label{sec-3.1}
The data show an emission line at $\sim 6.4$ keV in the rest frame of RBS 1124 (see Figure \ref{fig-2}). This feature is naturally interpreted as fluorescent emission from iron \citep{1991MNRAS.249..352G,2007MNRAS.382..194N}. Since the line is weak, we initially model it using a basic redshifted Gaussian model, \textit{zgauss}. We tried fitting the data in the 2 -- 10 keV energy band using the model \textit{tbabs$\times$(zgauss+powerlaw)}, leaving the width of the emission line free to vary. We could only find an upper limit for the line width, $\sigma < $ 0.3 keV. We therefore fixed the line width to $\sigma = 1$ eV, i.e. below our spectral resolution, assuming a narrow core iron line. This model combination gave a good fit with a $\chi^{2}$ of 981 for 996 d.o.f. and a line equivalent width of $42\pm17$ eV. This EW and $L_{\mathrm 2-10~\rm keV} = 6\times10^{44}$ erg s$^{-1}$ (\citealt{2010MNRAS.401.1315M}) agree with the EW-luminosity trend seen for AGN narrow iron emission lines due to the X-ray Baldwin effect (a known anti-correlation between the EW of the narrow emission line from neutral iron and 2 -- 10 keV X-ray luminosity of AGN, also known as the Iwasawa-Taniguchi effect; \citealt{1993ApJ...413L..15I,2007A&A...467L..19B}). We also tried adding a second, broad Gaussian line with a fixed line width in the source frame of 1 keV in order to test for the presence of a broad component. The inclusion of this broad Gaussian did not improve the fit. The upper limit for the EW of the broad line is estimated to be $\lesssim 75$ eV, which is significantly lower than the typical EWs of the broad iron line exhibited by unobscured AGN (EW $\sim 200$ eV; \citealt{2003ASPC..290...35R,2006AN....327.1032G}).

\subsection{Broadband Analysis}
\label{sec-3.2}

After focusing on the iron band specifically, we now explore whether the high-density relativistic reflection model offers a viable interpretation of the soft excess in RBS 1124. We use the Comptonization model \textit{nthcomp} (\citealt{1996MNRAS.283..193Z,1999MNRAS.309..561Z}) for the primary continuum of the source. The effects of cosmological redshift are convolved with the primary continuum using the \textit{zashift} model. The weak neutral iron line, which is likely a result of distant reflection from the torus, is fitted using the table model \textit{borus}\footnote{We use the borus12\_v190815a.fits file available at \url{https://sites.astro.caltech.edu/~mislavb/download/}} (\citealt{2019RNAAS...3..173B}). The \textit{borus} model assumes the reprocessing medium to have a spherical geometry with conical cutouts at the poles. Considering the bare nature of RBS 1124, we set up the \textit{borus} component in a way that its covering factor, cos($\theta _{\rm tor}$) is always less than the inclination angle, cos($\theta _{\rm inc}$). This ensures that the distant reflection is computed assuming that we do not look through the torus itself, but rather it is out of our line-of-sight. The photon index ($\Gamma$) and the electron temperature ($kT_{\rm e}$) values are linked between the \textit{nthcomp} and the \textit{borus} model components.

We try to model the disc reflection using two currently prominent reflection models: \textit{xillverCp} (\citealt{2010ApJ...718..695G,2013ApJ...768..146G}) and \textit{reflionx}\footnote{We use the reflionx\_HD\_nthcomp\_v2.fits file available at \url{https://www.michaelparker.space/reflionx-models}.} (\citealt{2005MNRAS.358..211R}). \textit{xillverCp} is an extension of \textit{xillver} - an ionized reflection table model (\citealt{2010ApJ...718..695G}) and is included in the RELXILL family of models (\citealt{2014ApJ...782...76G}). \textit{reflionx} is an ionized reflection table model developed by \cite{2005MNRAS.358..211R}. The latest versions of both models use \textit{nthcomp} as the incident primary continuum and allow the disc density to vary within the range $10^{15} - 10^{20}$ cm$^{-3}$. While using either model, the key parameters describing the ionising continuum, i.e., $\Gamma$ and $kT_{\rm e}$, are tied to the relevant parameters in the \textit{nthcomp} component. The other key reflection parameters, such as the iron abundance ($A_{\rm Fe}$)\footnote{We assume $(A_{\rm Fe})_{borus} = 0.7\times(A_{\rm Fe})_{xillver/reflionx}$, to account for the varying values assumed for the solar abundance of iron in \textit{borus} and \textit{xillver/reflionx} model components. 
}, ionisation parameter ($\xi$) and density ($n_{\rm e}$) of the disc are left free to vary. In order to apply the relativistic effects to the non-relativistic reflection spectrum, we use the convolution model \textit{relconv} (\citealt{2010MNRAS.409.1534D,2013MNRAS.430.1694D}). The \textit{relconv} model assumes the emissivity profile of the accretion disc to have a broken powerlaw in the form: $\epsilon (r) \propto r^{-q_{\rm in}}$  for $r_{\rm in} \leq r \leq r_{\rm br}$ and $\epsilon (r) \propto r^{-q_{\rm out}}$  for $r_{\rm br} \leq r \leq r_{\rm out}$, where $q_{\rm in}$ and $ q_{\rm out}$ denote the inner emissivity index and outer emissivity index, respectively. $r_{\rm in}, r_{\rm out}$ and $r_{\rm br}$  are the inner radius, outer radius and the break radius of the accretion disc. We allow $r_{\rm br}$, $q_{\rm in}$, the spin parameter ($a^*$) and the inclination angle (\textit{i}) to vary, while we assume that for $r~ >~ r_{\rm br}$ the emissivity can be well approximated by the Newtonian limit, and thus adopt a fixed value of $ q_{\rm out} = 3$ for the outer emissivity index (\citealt{1997ApJ...488..109R,2012MNRAS.424..217F}). We therefore have two resultant models to describe the relativistic reflection component within the broadband spectrum: \textit{relconv$\otimes$xillverCp} (hereafter referred to as Model 1) and \textit{relconv$\otimes$reflionx} (Model 2). It is worth noting that  Model 1 can essentially be replaced by \textit{relxillCp}, which is also a part of the broad RELXILL model ensemble \citep{2013ApJ...768..146G}. However, such a substitution cannot be extended to \textit{reflionx}. To maintain consistency between the two models considered here, we opt to use \textit{relconv$\otimes$xillverCp/reflionx}.

\begin{figure*}
  \centering
    \begin{minipage}[b]{0.35\textwidth}
  \hspace*{-1.4cm}
   \hspace*{1cm}
    \includegraphics[scale=0.3]{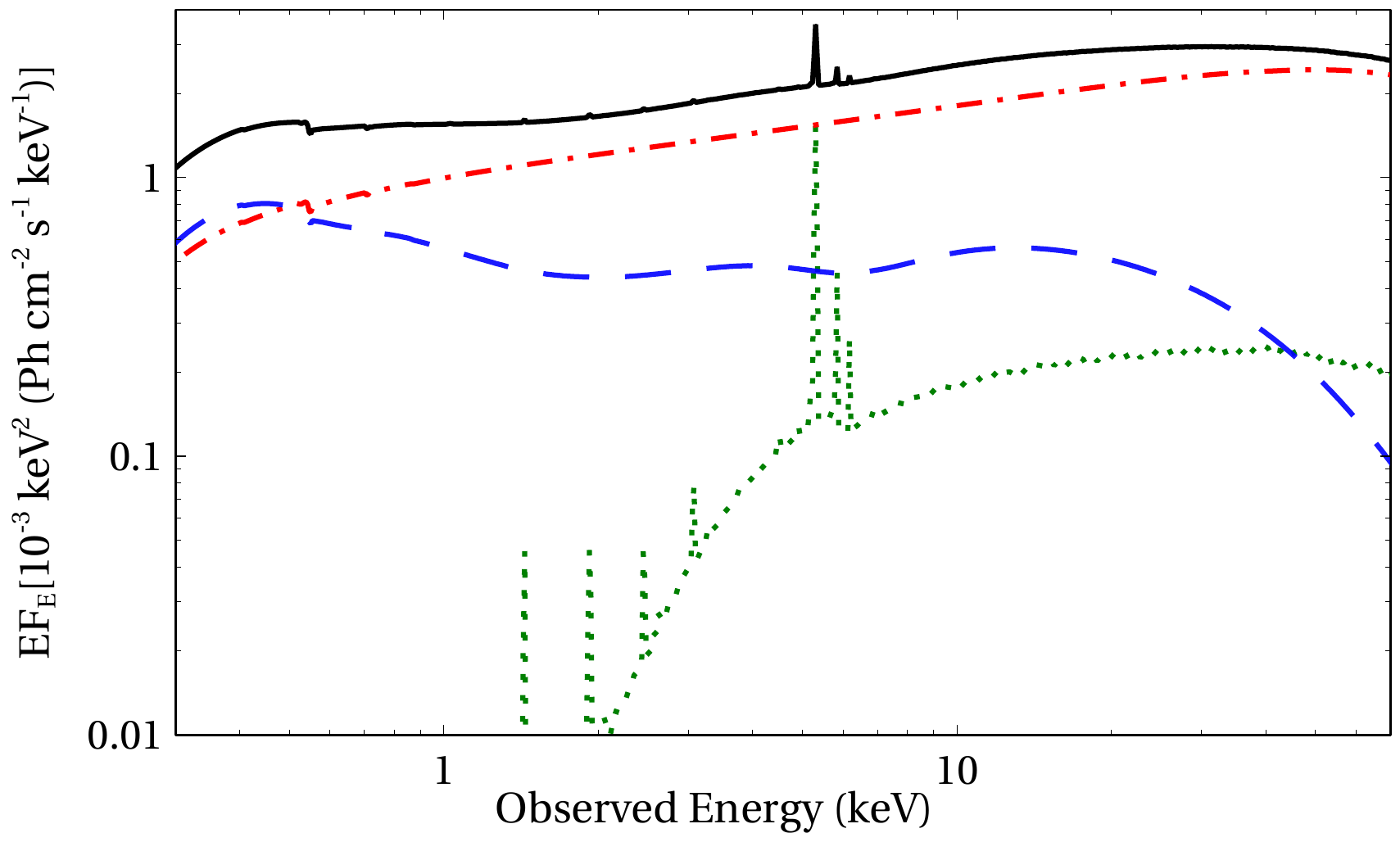}
     \label{fig-4a}
  \end{minipage}
  \hfill
  \begin{minipage}[b]{0.35\textwidth}
  \hspace*{-2.35cm}
    \includegraphics[scale=0.3]{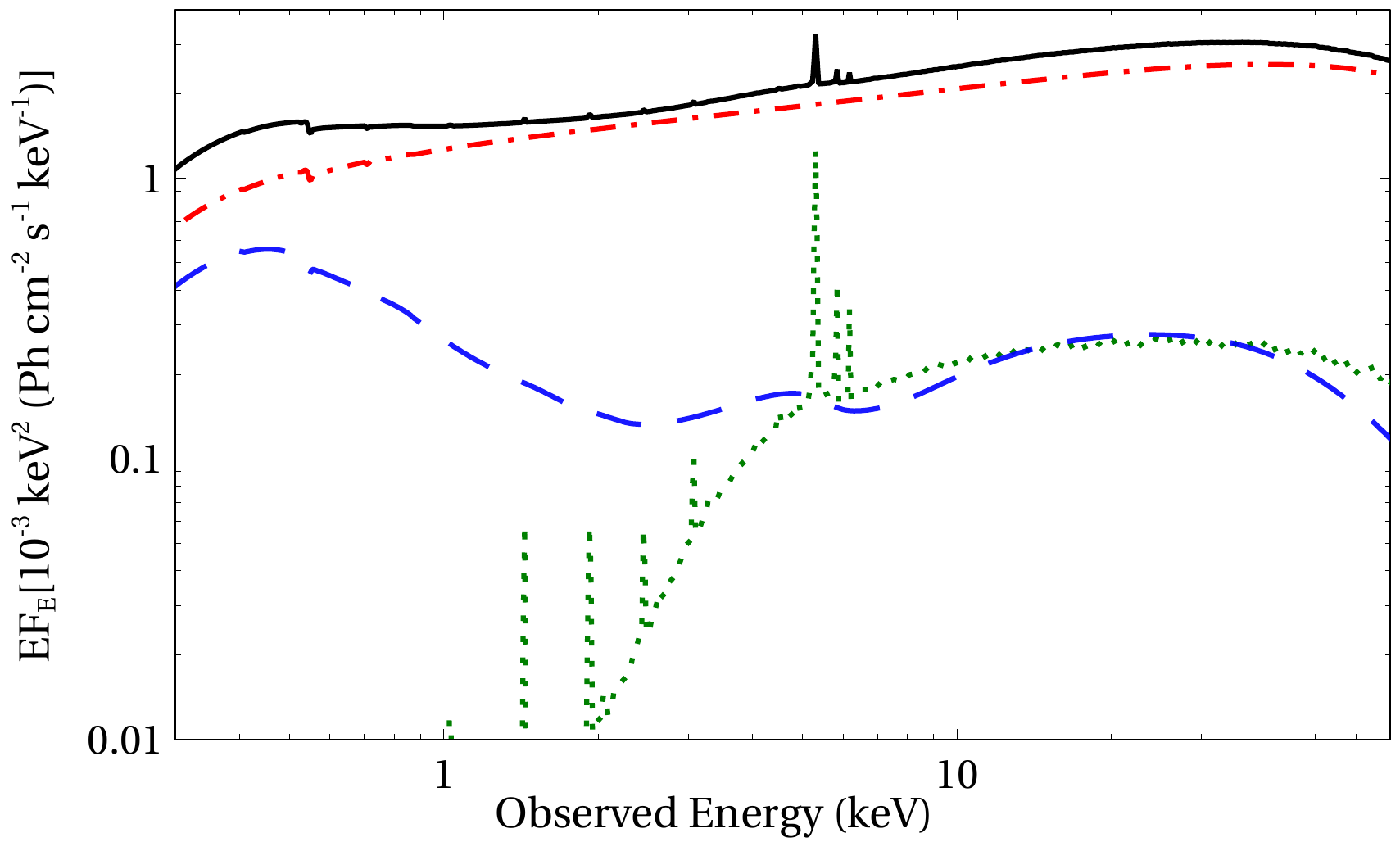}
    \label{fig-4b}
  \end{minipage}
  \caption{The relative contributions from various model components from Model 1 (left) and Model 2 (right). Red dash-dotted lines denotes the Comptonised continuum, green dotted lines denote the distant reflection, and blue dashed lines denote the relativistic reflection. \label{fig-4}}
\end{figure*}

\begin{figure}
\centering
\hspace*{-1cm}
\includegraphics[scale=0.38]{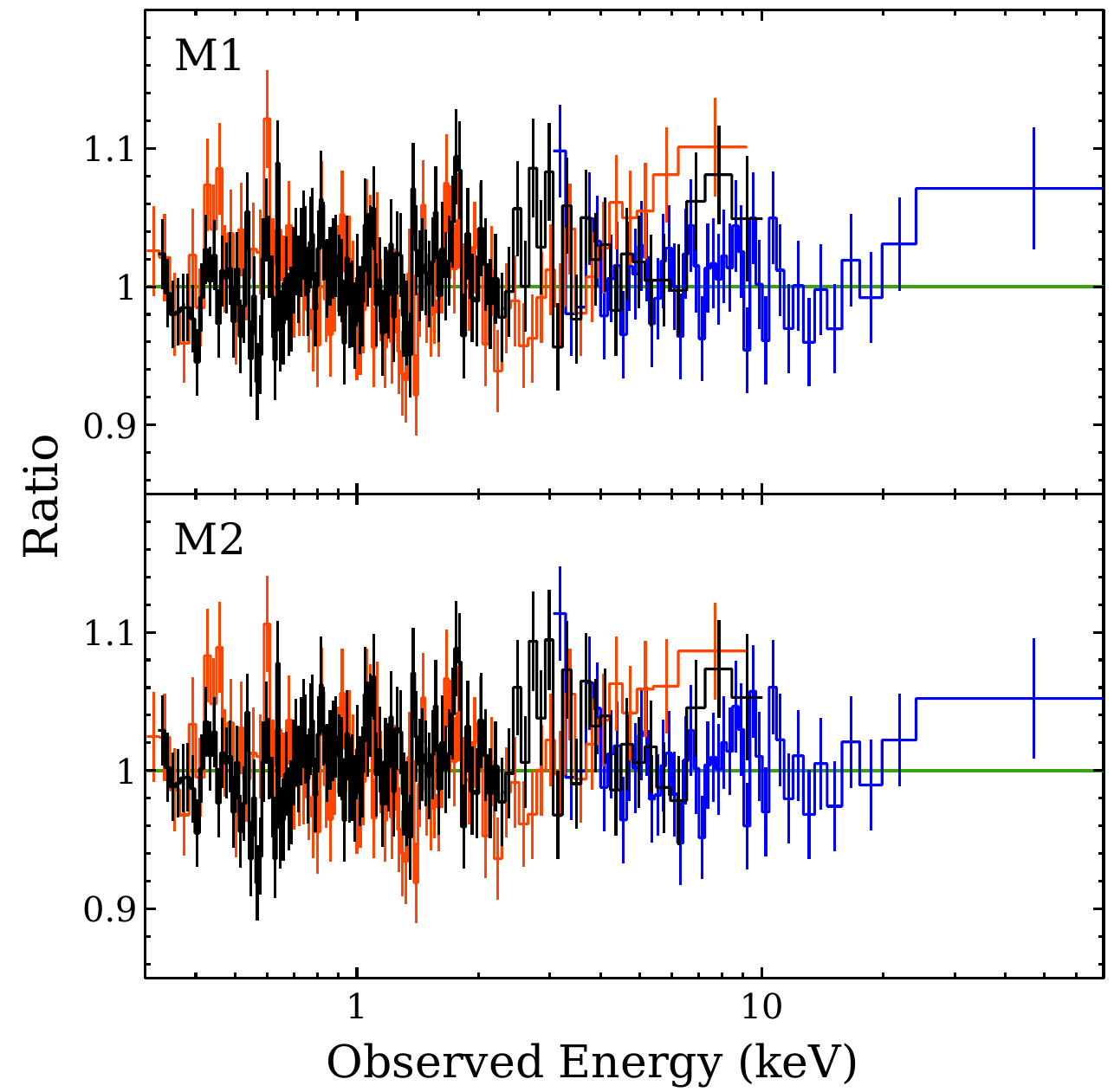}
\caption{Data-to-model ratio for the best-fit obtained from Model 1 (top) and Model 2 (bottom). Note that the colours have the same meanings as in Figure \ref{fig-2}.}
\label{fig-5}
\end{figure}

Model 1, along with the \textit{tbabs, nthcomp} and \textit{borus} components discussed above, gives an excellent fit with $\chi^{2}/d.o.f.$ of 1913.9/1894. The relative contributions from the different model components for the best-fit parameter combination in Model 1 are shown in the left panel of Figure \ref{fig-4} and the best-fit parameter constraints are given in Table \ref{tab 2}. The model fits all the features in the broadband energy range relatively well. However the data at higher energies slightly deviate from the fit. The data-to-model ratio plot for Model 1 is shown in panel M1 of Figure \ref{fig-5}. Model 2 gives a slightly improved fit with a $\Delta\chi^{2}$ of -3 and also fits the data at higher energies (see panel M2 in Figure \ref{fig-5}). The relative contributions for the best-fit parameters in Model 2 and these parameter constraints are again presented in Figure \ref{fig-4} and Table \ref{tab 2}, respectively. The results obtained from each fit are discussed in Section \ref{sec-4}.

Along with parameters directly included in the models used here, we also compute the reflection fraction ($R = \Omega/2\pi$) for both our model fits following the methodology described in \cite{2013MNRAS.428.2901W}. The reflection fraction provides insights into the geometry of the disc-corona system. Reflection fraction ($R$) conventionally refers to the ratio of the reflected flux ($F_{\rm Ref}$) to the Comptonized flux ($F_{\rm Com}$). For a semi-infinite slab in an ideal non-relativistic scenario, the accretion disc covers a solid angle ($\Omega$) of 2$\pi$ steradians as seen from the corona, which is traditionally defined as a reflection fraction of 1. To estimate the reflection fractions implied by our fits, we calculate the flux ratios between the reflection and the primary continuum in the 15--50\,keV band (as this energy band is dominated by the Compton scattering processes and is not strongly influenced by the ionisation state of the disc) prior to the application of any relativistic blurring, and then compare this to the equivalent flux ratio predicted by the \textit{xillverCp} model 
for $R = 1$ for the same set of key spectral parameters ($\Gamma$, $i$, $\xi$ and $A_{\rm Fe}$). These values are also reported in Table \ref{tab 2}. Note that even though they are based on a flux ratio, since they are scaled to the $R=1$ case from \textit{xillverCp} they should be broadly analogous to the definition of the reflection fraction discussed in \cite{2016A&A...590A..76D}, as opposed to the reflection `strength' also discussed in that work (which is a flux ratio), even if our definitions are not exactly the same.

We also perform Markov chain Monte Carlo (MCMC) analysis to explore the parameter space to get robust constraints on parameters. The \textsc{chain} command within \textsc{xspec} was invoked to do the  MCMC analysis. We use the Goodman-Weare (GW) algorithm with 60 walkers and a chain length of 30,000 for each walker and a burn length of 5000, thus giving a total chain length of 1,500,000. Errors for the parameters quoted in Table \ref{tab 2} are estimated based on these chain values. The integrated posterior distribution for each parameter is as shown in Figures \ref{fig-A1} and \ref{fig-A2} (in Appendix \ref{Appendix}) for Models 1 and 2, respectively. Error estimates for several parameters are consistent across both the models.

\section{Results}
\label{sec-4}
 
Models 1 and 2 produce equally good fits with $\chi^{2}$/d.o.f. $~ \sim 1.01$ and the parameters obtained from both the models are broadly consistent (see Table \ref{tab 2}). $\Gamma$ values of  $1.76^{+0.03}_{-0.02}$ and $1.80^{+0.02}_{-0.01}$, obtained from Models 1 and 2 respectively, are in fair agreement with the $\Gamma$ value obtained by previous works ($\Gamma \sim 1.85$; \citealt{2010MNRAS.401.1315M,2013MNRAS.428.2901W}). Both models predict a high density disc for RBS 1124 with log($n_{e}$/cm$^{-3}) > ~19.2$. While Model 1 could only achieve a lower limit to the density value, Model 2 provides two-sided limits on the density with log($n_{e}$/cm$^{-3}) = 19.48^{+0.28}_{-0.31}$. We observe that, for a given density value, the \textit{reflionx} model generates higher flux for the reflected continuum at energies below 1 keV compared to the \textit{xillverCp} model, which allows it to effectively account for the soft excess at slightly lower densities within the range covered by these models. Consequently, its lower best-fit density allows \textit{reflionx} to provide two-sided constraints, while its higher best fit density means the \textit{xillverCp} confidence limits run up against the current upper limit of the model.

\begin{table}
\centering
\caption{Best-fit parameters obtained from the broadband analysis of \textit{XMM-Newton+NuSTAR} data of RBS 1124 using Models 1 and 2. All errors are quoted at 90\% confidence level.}
\begin{tabular}{llcc}
\hline 
Model & Parameters & 1 & 2  \\
\hline
\textit{nthComp} & $\Gamma$ & $1.76^{+0.03}_{-0.02}$ &  $1.80^{+0.02}_{-0.01}$ \\
& $kT_{\rm e}$ (keV) & $>~41$ &     $>~34$  \\
& Norm ($\times10^{-3}$) & $1.40^{+0.30}_{-0.10}$ &  $1.75^{+0.18}_{-0.01}$ \\
\\
\textit{borus} & log ($N_{\rm H}$ cm$^{-2}$) & $23.19^{+0.29}_{-0.16}$  &  $23.12^{+0.24}_{-0.14}$ \\
& C factor & $0.92_{-0.29}$ &  $0.93_{-0.28}$ \\
\\
\textit{relconv} & $q_{\rm in}$ & $5.01^{+2.49}_{-0.16}$ &    $>~3.94$  \\
& $R_{\rm br}$ ($R_{\rm ISCO}$) & $>~4.85$  &  $>~3.01$  \\
& $a^{*}$ & $>~0.865$  &  $>~0.236$ \\
& \textit{i} ($\deg$) & $<~42.84$  &  $<~39.41 $ \\
\\
\textit{xillverCp} & log ($\xi$ erg cm s$^{-1}$)  & $3.11^{+0.06}_{-0.13}$    & -  \\
& log ($n_{\rm e}$/cm$^{-3}$) & $>~19.85$ & - \\
& A$_{\rm Fe}$ (solar) & $0.93^{+0.39}_{-0.20}$  &  -  \\
& Norm ($\times 10^{-5}$) & $1.17^{+0.42}_{-0.40}$  &  -   \\
\\
\textit{reflionx} & log ($\xi$ erg cm s$^{-1}$)  & - &  $2.85^{+0.13}_{-0.16}$   \\
& log ($n_{\rm e}$/cm$^{-3}$) & - & $19.48^{+0.28}_{-0.31}$  \\
& A$_{\rm Fe}$ (solar) & -  &  $0.62_{-0.11}^{+0.12}$   \\
& Norm ($\times 10^{-2}$) & - & $1.78^{+0.02}_{-0.70}$ \\

\hline
\multicolumn{2}{|c|}{Reflection fraction (R)}  & $0.44^{+0.33}_{-0.21}$  & $0.17^{+0.20}_{-0.12}$ \\
\hline
\multicolumn{2}{|c|}{$\chi^{2}$/d.o.f.} & 1913.9/1894 & 1909.7/1894   \\
\hline 
\end{tabular}
\begin{flushleft}
 \end{flushleft}
\label{tab 2}
\end{table}

\begin{figure}
\centering
\hspace*{-0.15cm}
\includegraphics[scale=0.33]{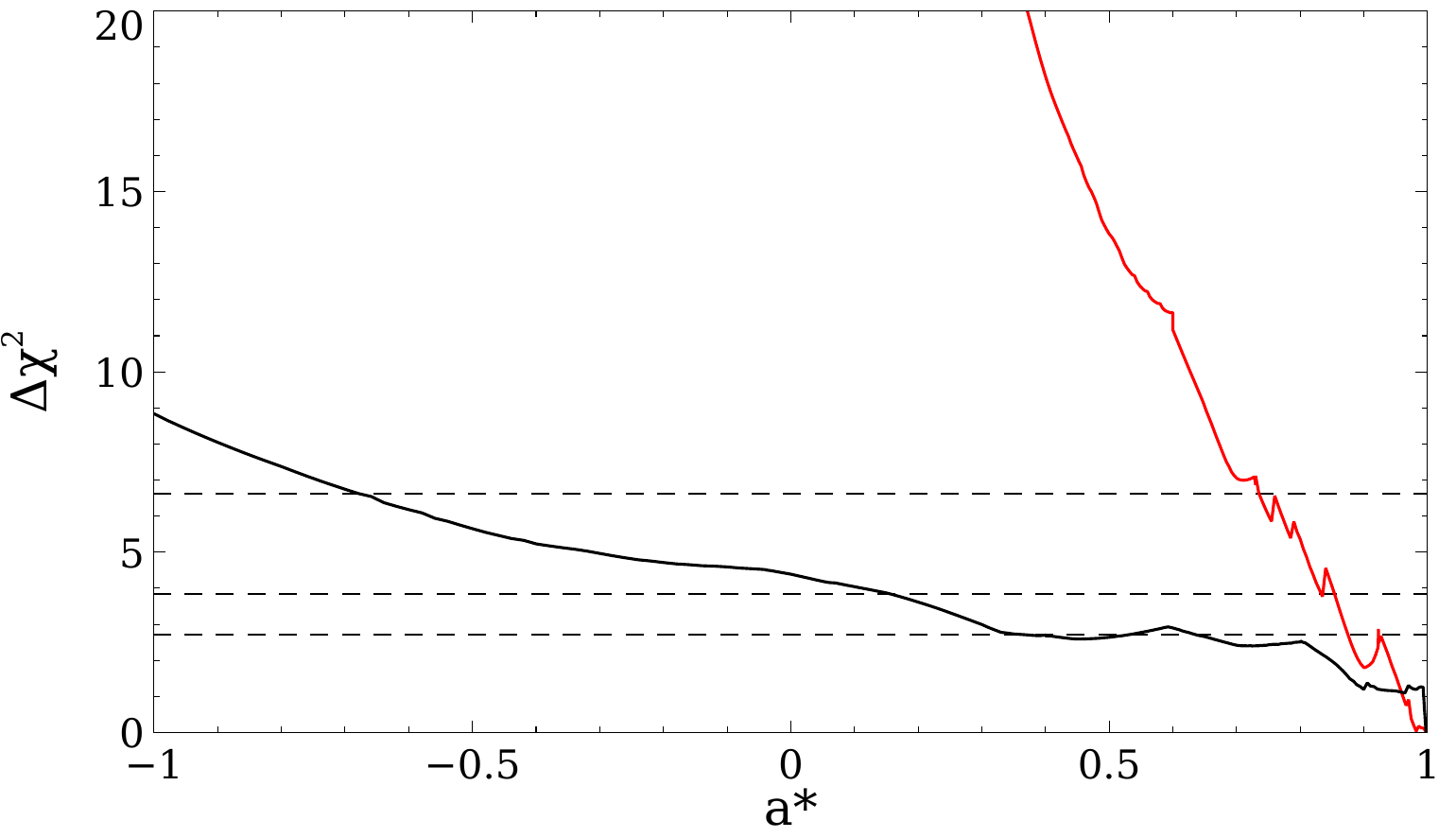}
\caption{Variation of fit-statistic by stepping the spin parameter $a^{*}$ away from the best fit for both Model 1 (red) and Model 2 (black).
The dotted lines signify the 90\%, 95\%, and 99\% confidence intervals, respectively. \label{fig-6}}

\end{figure}

Model 1 prefers a highly spinning black hole ($a^{*} > 0.865$), whereas Model 2 presents notably weaker constraints on the spin ($a^{*} > 0.236$). Variation in $\chi^2$ as a function of $a^{*}$, obtained by stepping $a^{*}$ using the \textsc{steppar} command in \textsc{xspec}, for Models 1 and 2 are presented in the Figure \ref{fig-6}. This is related to the same issue as highlighted above; the stronger reflection continuum at low energies in \textit{reflionx} means this model does not need to rely so heavily on the line emission to reproduce the overall soft X-ray flux. In contrast, \textit{xillverCp} shows a stronger need for extreme line broadening in order to produce a smooth soft excess, as the lines make a larger relative contribution to the soft excess. The different relative levels of the reflection continuum at low energies in the two models relates to differences in the fine details of the calculations performed, e.g. the atomic physics included, the treatment of scattering, and the treatment of resonant emission lines. These factors dictate the level of photoionised absorption, the strength of key emission lines and the energy at which the low-energy continuum from free-free emission peaks, all of which can impact the flux of the reflection model at the lowest energies in the observed bandpass in particular.

Model 1 yields an iron abundance ($A_{\rm Fe}$) in the range 0.7 -- 1.2, consistent with the solar value, while Model 2 shows an abundance value $< 0.7$. Both are lower than the constraint obtained previously ($A_{\rm Fe} = 2.7^{+1.8}_{-0.9}$; \citealt{2013MNRAS.428.2901W}). This difference may be attributed to the implementation of a variable-disc density model in this work. According to discussions in \cite{2018ApJ...855....3T} and \cite{2019MNRAS.489.3436J}, the high-density disc model leads to an increase in the continuum within the reflection component around the iron line, thus giving lower iron abundances when compared to the results obtained fitting older models in which the density was fixed to $10^{15}$ $\rm cm^{-3}$. The emissivity index could not be well constrained in our analysis. A lower limit of 3.94 was obtained from Model 2. The break radius is also poorly constrained, with $R_{\rm br} > 3.01R_{\rm ISCO}$. High ionization values were obtained from our fit, while the previous analyses of the source find lower values for ionization ($\xi \leq 100$ erg cm s$^{-1}$; \citealt{2010MNRAS.401.1315M,2013MNRAS.428.2901W,2019ApJ...874..135T}). Our reflection modelling implies that the source is viewed at a low inclination ($i < 43^{\circ}$), which is in line with broad expectations for the unified model given the bare nature of RBS 1124.

\section{Discussion}
\label{sec-5}

We present a broadband (0.3 -- 70 keV) spectral analysis of the broad line Sy 1 AGN RBS 1124, using a new co-ordinated \textit{XMM-Newton} and \textit{NuSTAR} observation. The spectrum exhibits features typically observed in `bare' AGN: a soft excess and weak distant reflection along with the primary X-ray continuum. Our work focuses on investigating the relativistic reflection model for the soft excess seen in this source, given the ongoing debate in the literature regarding its ability to explain the broadband datasets available for bare AGN \citep{2019ApJ...871...88G,2020MNRAS.497.2352M,2021ApJ...913...13X,2021A&A...654A..89P}. We emphasise that the warm corona model remains a well-established alternative for the soft excess. However, there is not a corresponding debate, as this model is invariably capable of reproducing the broadband spectrum for bare AGN, since the soft and hard X-ray components are essentially de-coupled from a fitting perspective. In this context, we concentrate on testing the compatibility of the high-density reflection model with the broadband data for RBS 1124, focusing on whether this model can reproduce the full spectral range. 
This study marks the first application of variable-density relativistic reflection models to RBS 1124, and we test both of the reflection models commonly used in the recent literature (\textit{xillver} and \textit{reflionx}). As discussed in more detail below, we find that high-density relativistic reflection model can indeed reproduce the broadband spectrum seen from this bare AGN.

Previous studies of this source used fixed density relativistic reflection models ($n_{e} = 10^{15}$ cm$^{-3}$) to describe the broadband spectrum (\citealt{2009MNRAS.394..443M,2013MNRAS.428.2901W,2019ApJ...874..135T}). These models solely relied on the relativistic blurring of emission lines to explain the soft excess, resulting in tighter constraints on earlier spin measurements. Our analysis yields spin constraints that, while formally weaker than these previous relativistic blurring studies, are likely more realistic owing to the higher S/N data over a broader bandpass and the inclusion of disc density as a free parameter in our models. Similarly, prior studies inferred varying reflection fractions using low S/N \textit{Suzaku} PIN data: \cite{2013MNRAS.428.2901W} inferred a large reflection fraction for RBS 1124 ($R \sim$ 3), while \cite{2009MNRAS.394..443M} found a much lower value ($R \simeq 0.4$) from the same data. Our results with the much better \textit{NuSTAR} high-energy data suggest the latter interpretation is more reliable.

\subsection{General Properties of the Primary Continuum}
\label{sec-5.1}
 
The corona around an SMBH is characterized by two key parameters: photon index ($\Gamma$) and high energy cut-off ($E_{\rm cut}$). $E_{\rm cut}$ is directly associated with the  plasma temperature ($kT_{\rm e}$) by the 
 relation  $E_{\rm cut} = (2-3)kT_{\rm e}$ (\citealt{2001ApJ...556..716P}). 
The parameters $\Gamma$ and $kT_{\rm e}$ are directly estimated from the spectral fitting through the Comptonizing model component. Our best-fit model indicates a $\Gamma = 1.80\pm0.01$ for the primary continuum of RBS 1124, which agrees well with the typical value of $1.80\pm0.02$ seen in unobscured AGN  (\citealt{2017ApJS..233...17R}; see also \citealt{1994MNRAS.268..405N}). We obtain a lower limit on $kT_{\rm e}$ ($ > 34$ keV) from our best-fit. Note that  although $\Gamma$ had been extensively monitored previously for this source (see Section \ref{sec-4}), $kT_{\rm e}$ could not be previously constrained due to limitations on coverage and data quality in the high energy band-pass prior to \textit{NuSTAR}. The relation between $E_{\rm cut}$ for the Comptonizing plasma and Eddington ratio ($\lambda_{\rm Edd}$) was investigated by \cite{2018MNRAS.480.1819R} by studying a sample of 317 unobscured AGN. Their results show a median value of $E_{\rm cut}=160\pm41$ keV for the sample for sources with $\lambda_{\rm Edd} > 0.1$. For a bolometric luminosity ($L_{\rm Bol}$) of $3.4\times10^{45}$ erg s$^{-1}$ (\citealt{2004AJ....127..156G}) and an Eddington luminosity ($L_{\rm Edd}$) of $2.3\times10^{46}$ erg s$^{-1}$ given the mass $1.8\times10^{8}M_{\odot}$ (\citealt{2010MNRAS.401.1315M}), RBS 1124 is inferred to be accreting with $\lambda_{\rm Edd}$ of 0.145. Our $E_{\rm cut}$ and $\lambda_{\rm Edd}$ values are consistent with \cite{2018MNRAS.480.1819R}. These results indicate that RBS 1124 shows the typical plasma characteristics exhibited by unobscured AGN.

\subsection{Soft Excess and the High Disc Density of RBS 1124}
\label{sec-5.2}

The soft excess has consistently appeared as a prominent characteristic in the X-ray spectra of unobscured Sy1 AGN accreting with $\lambda_{\rm Edd} \geq 0.01$, yet its true origin remains unclear. Currently, two leading models, namely the the warm corona model and the relativistic reflection model, try to explain the origin of the soft excess. Both models adopt distinct physical assumptions about the  origin of this emission, with diverse insights into the properties of the inner accretion flow. The warm corona model invokes a radially stratified accretion flow, and thus potentially provides insight into how this stratification occurs (\citealt{2013A&A...549A..73P,2018MNRAS.480.1247K}). In contrast, the relativistic reflection interpretation invokes reprocessing of the hard X-ray radiation by the innermost accretion flow, and potentially provides insight into the ionisation state and composition of the accretion flow as well as the geometry of the illuminating X-ray source. The high-density models are a recent evolution of this latter interpretation (\citealt{2013ApJ...768..146G,2013MNRAS.428.2901W,2022MNRAS.513.4361M}).

Reflection from the inner disc also gives information about the spin of the black hole, by assessing the extent of the relativistic effects that broaden the observed reflection spectrum (\citealt{1996rftu.proc..455I,2006MNRAS.365.1067C,2013MNRAS.428.2901W,2021ARA&A..59..117R}). The spin measurements of supermassive black holes (SMBHs) serve as important tracers of their growth history, as their formation and evolutionary trajectories are reflected on the mass and spin of the SMBHs \citep{2023arXiv231104752P}. For instance, SMBHs formed via steady radiatively-efficient accretion are expected to exhibit rapid rotation. On the other hand, SMBHs formed via mergers are expected to exhibit moderate spin values, as the merger events typically reduce spin of the black hole (\citealt{2003ApJ...585L.101H,2005ApJ...620...69V,2020ApJ...895...95P}). Therefore, a comprehensive mass-spin distribution of SMBHs would help distinguish the different evolutionary paths they have followed. 
Moreover, spin of the black hole may influence its efficiency in generating powerful relativistic jets (\citealt{1977MNRAS.179..433B}) making robust spin measurements all the more crucial. Studying the relativistic reflection from the innermost accretion disc is the leading method for measuring black hole spin in active galaxies, and so it is critical to assess the degree to which the soft excess is related to this emission.

A key part of determining the origin of the soft excess is testing these models against the best data available. For the reflection model, this requires broadband spectroscopy as soft excess is assumed to be part of the broader reflection spectrum, which has key features in the iron band and in the hard X-ray band. We focus on testing this model using our new broadband dataset on the bare AGN RBS\,1124 along with the very latest variable-density reflection models. It is worth noting again that there are still only limited instances where high S/N broadband spectra are available for the bare AGN that are ideally suited to such observational tests, and that there are various conflicting claims in the literature about whether the reflection model provides acceptable fits to the data in these cases (\citealt{2018A&A...609A..42P,2019ApJ...871...88G}). We find that the reflection model does indeed fit the broadband data well for RBS 1124, and seems to give broadly sensible results in terms of the primary continuum properties (see Section \ref{sec-3.2}) and the inferred inclination of $i < 43^{\circ}$ is as expected for a bare AGN based on the unified model (\citealt{1993ARA&A..31..473A}), both of which are useful sanity checks.

Another key characteristic of the more recent evolution of the relativistic reflection model is that it offers a rare opportunity to place observational constraints on the density of the inner disc \citep{2019MNRAS.489.3436J,2022MNRAS.513.4361M}. The density of a radiation pressure dominated accretion disc, as described by Equation 8 in the disc-corona model of \cite{1994ApJ...436..599S}, is given by 
\begin{equation}
   n_e = \frac{1}{\sigma_{T}}\frac{256\sqrt{2}}{27}\alpha^{-1}R^{3/2}\dot{m}^{-2}[1-(R_{\rm in}/R)^{1/2}]^{-2}\xi^\prime(1-f)^{-3},
   \label{eq-1}
\end{equation}
where $\sigma_{T} = 6.64\times10^{-25}$ cm$^{2}$ is the Thomson cross-section; $\dot{m}$ is the dimensionless accretion rate expressed in terms of the Eddington luminosity ($\dot{m} = \dot{M}c^{2}L_{\rm Edd}^{-1}$), $R_{\rm in}$ is the inner radius of the disc and is set to $R_{\rm ISCO}$ of a maximally spinning black hole; $R_{\rm s}$ is the Schwarzschild radius; \textit{R} is the disc radius in units of  $R_{\rm s}$; $\alpha$ is the viscosity parameter of the disc and is set to 0.1; $\xi^\prime$ is the conversion factor in the radiative diffusion equation and chosen to be 1 and \textit{f} is the ratio of power dissipated to the corona from the disc varies between 0 and 1 (reducing to the standard thin disc solution of \citealt{1973A&A....24..337S} in the former limit). 
As a key parameter in this disc-corona model, accurate determinations of density play a key role in determining its viability. We cannot statistically distinguish between the \textit{reflionx} and \textit{xillver} reflection models, but the combined implication of our analysis is that RBS 1124 has an inner disc density log ($n_{\rm e}$/cm$^{-3}) \geq 19.2$.

Among the relatively small sample of other AGN where variable-density reflection models have been applied (\citealt{2019MNRAS.489.3436J,2022MNRAS.513.4361M}), a broad anti-correlation has been observed between the density ($n_{\rm e}$) and the mass of the SMBH ($m_{\rm BH}$). RBS 1124, hosting a SMBH with a mass of $1.8\times10^{8}M_{\odot}$, initially seems to stand apart from other Sy 1 AGN as our results suggest it has a high density for the inner disc with log(n$_{\rm e}$/cm$^{-3}) > 19.2$. Nevertheless, for a radiation pressure dominated accretion disc, the density of the disc is expected to be anti-correlated to both $m_{\rm BH}$ and $\dot{m}^{2}$, by the relation log($n_{\rm e}) \propto - $ log$(m_{\rm BH}\dot{m}^{2}$), based on Equation \ref{eq-1}.

\begin{figure}
\centering
\includegraphics[scale=0.39]{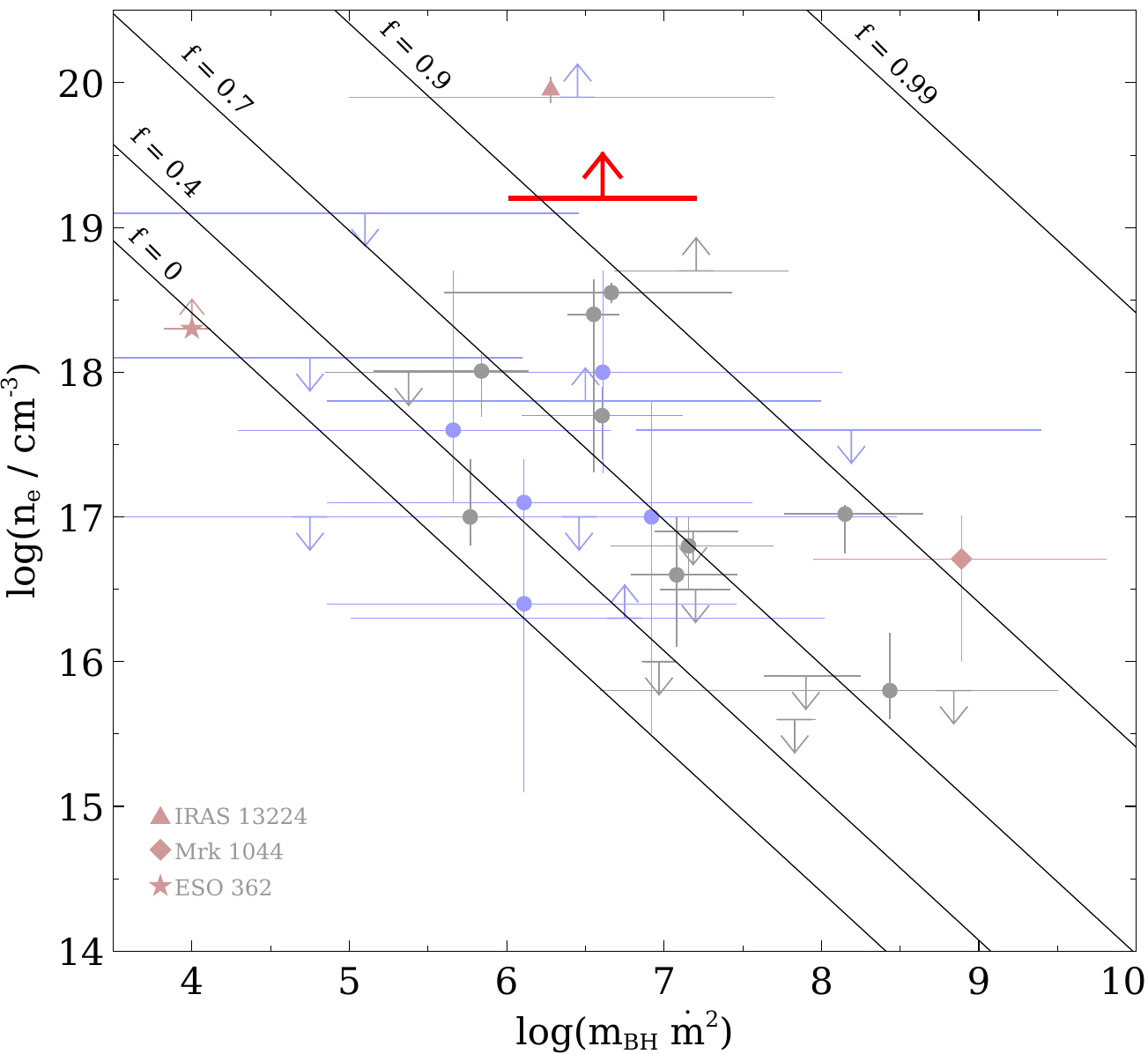}
\caption{Variation of the disc density log($n_{\rm e}$/cm$^{-3}$) with respect to the product of mass of the black hole ($m_{\rm BH}$) and $\dot{m}^2$ (where $\dot{m} = \dot{M}c^{2}L_{\rm Edd}^{-1}$ is the dimensionless accretion rate expressed in terms of the Eddington luminosity). The points in the background correspond to a combined sample of 17 Sy 1 AGN \citep{2019MNRAS.489.3436J} and 13 low mass AGN from \citep{2022MNRAS.513.4361M}, plotted in black and blue points, respectively. Sy 1 IRAS 13224 \citep{2022MNRAS.514.1107J},  Sy 1 AGN Mrk 1044 \citep{2018MNRAS.479..615M} and ESO 362-G18 \citep{2021ApJ...913...13X} are additionally included in the background sample as triangle, diamond and star shaped data points, respectively. Upward and downward arrows indicate that only lower and upper density limits could be derived from the analysis. The upward arrow in red, highlighted in the foreground, shows the position of RBS 1124 in the diagram. Black lines denote solutions of the density of the radiation pressure dominated accretion disc for \textit{f} values of $0, 0.4, 0.7, 0.9$ and $0.99$. See text for details.}
\label{fig-7}
\end{figure}

In case of RBS 1124, for $\dot{m} = 0.145$, we estimate log($m_{\rm BH}\dot{m}^{2}) = 6.57\pm0.6~\rm dex$\footnote{We assume that the uncertainty here is dominated by uncertainty in the black hole mass. The quoted uncertainty therefore combines in quadrature the 0.4 dex scatter on the scaling relation between $m_{\rm BH}$, $L_{5100}$ and H$\beta$ line width (\citealt{2006ApJ...641..689V}) used to estimate the mass of RBS 1124, and the 0.4 dex uncertainty on the reverberation mapping results used to calibrate this scaling relation (\citealt{2014SSRv..183..253P}).}. Figure \ref{fig-7} shows log(n$_{\rm e}$/cm$^{-3})$ versus log($m_{\rm BH}\dot{m}^{2}$) for a combined sample of Type 1 AGN from \cite{2018MNRAS.479..615M}, \cite{2019MNRAS.489.3436J}, \cite{2021ApJ...913...13X}, \cite{2022MNRAS.513.4361M} and \cite{2022MNRAS.514.1107J}. 
The red arrow in the figure represents the constraints we can place for RBS 1124, indicating its position relative to the rest of the Sy 1 AGN shown. 
Although the density of RBS 1124 obtained from our analysis is relatively high compared to other AGN, it still generally adheres to the anti-correlation proposed by \cite{1994ApJ...436..599S}, considering the wide scatter observed in other AGN. The solid lines in the figure represent solutions for the density of a radiation pressure dominated disc (Equation \ref{eq-1}) for different values of $f$. Given the position of RBS 1124 in the log(n$_{\rm e}$/cm$^{-3})$ versus log($m_{\rm BH}\dot{m}^{2}$) plane, it can be inferred that more than 90\% of the power is dissipated to the corona from the disc. While the coronal dissipation fraction in RBS 1124 seems high compared to the \cite{2019MNRAS.489.3436J} sample, such high values have been seen previously.  IRAS 13224-3809 \citep{2022MNRAS.514.1107J} and Mrk 509 \citep{2019ApJ...871...88G} both exhibit dissipation fraction values exceeding 0.85. A similar study on a group of high-mass NLSy1s at high redshifts revealed that more than 95\% of the power seems to be dissipated in the corona in the case of E1346+266 \citep{2023MNRAS.522.5456Y}.

\subsection{Reflection Strength}
\label{sec-5.3}

The reflection strength gives information about the geometry of the corona and the proximity of the inner edge of the accretion disc to the central black hole. A compact corona  located right above the event horizon of a rapidly spinning SMBH can exhibit high reflection strengths, due to the effects of light bending \citep{2004MNRAS.349.1435M,2012MNRAS.419..116F,2014MNRAS.443.1723P,2014MNRAS.444L.100D,2016A&A...590A..76D,2021MNRAS.506.1557W}. The reflection strength however tends to converge to the standard non-relativistic case ($R = 1$), for sources with low spin or when the compact corona is located at a larger height \citep{2016A&A...590A..76D}. The impact of the black hole spin and height of the corona on the reflection strength has been extensively studied for a lamp-post model of the corona, which is likely a simplified representation of the real coronal geometry for many AGN. In our work, we do not prescribe any specific coronal geometry. Instead, we assume a broken power-law emissivity profile for the reflected spectrum, a profile that can be produced by  a range of coronal geometries.

Although the spin and emissivity index values could not be tightly constrained, our fits did establish lower limits for both parameters. Our analysis suggests a moderately spinning black hole with $a^{*} > 0.236$ and an emissivity index $q_1 \gtrsim 3.94$. These values typically correspond to standard reflection strength $\sim$ 1 \citep{1995MNRAS.273..837M}. However, our estimates show reflection strength values of 0.4 and 0.2 for Models 1 and 2, respectively, consistent with the findings of \cite{2010MNRAS.401.1315M} who estimated an R value of 0.4. These values, significantly less than 1, could be indicative of an outflowing corona \citep{1999ApJ...510L.123B}, where corona is accelerated away from the disc at mildly relativistic velocities by radiation pressure \citep{2007MNRAS.377.1375G,2020MNRAS.496.3708G,2022MNRAS.512..761W}. Consequently, coronal X-rays are preferentially emitted upward, away from the disc, resulting in direct observation as continuum emission, leading to diminished illumination of the disc and a reduced reflection strength.

\subsection{Absence of observable wind signatures}
\label{sec-5.4}

Ionized wide-angle outflows are ubiquitous features in accretion-powered sources \citep{1982MNRAS.199..883B,2000ApJ...543..686P,2004ApJ...616..688P,2019A&A...630A..94G}. AGN accreting at moderate - high accretion rates ($\dot{m} > 0.1$) are expected in particular to launch radiation-driven accretion disc winds \citep{2004ApJ...616..688P,2010A&A...516A..89R}. These winds manifest as blueshifted absorption lines in the spectrum of X-ray sources \citep{2004ApJ...609..325U,2006Natur.441..953M,2006AN....327.1012C,2009ApJ...694....1S} and are commonly observed in AGN \citep{1997MNRAS.286..513R,2003ARA&A..41..117C,2010A&A...521A..57T,2014MNRAS.441.2613L}.

The absence of detectable wind signatures in RBS 1124, despite its moderately high mass accretion rate ($\dot{m} \sim 0.14$), is somewhat unexpected. While we cannot definitively conclude the absence of winds based solely on X-ray spectroscopy, we can propose plausible explanations for their absence in the X-ray spectrum. One plausible explanation is the viewing angle, which in the case of RBS 1124 is close to face-on, consistent with its Type 1 classification as per the unification theory. In Type 1 systems with equatorial winds, our line of sight may not directly intercept with the winds. However, there could be other sight lines where the winds are observable. Optical spectroscopic studies of the source could offer further insights into the nature of the outflows in the system. For instance, RBS 1055 is one of the brightest radio-quiet AGN detected in the ROSAT survey, with a luminosity of log$(L_{(\mathrm {0.5 - 2 \rm ~keV})}/[\rm {erg~s^{-1}}])$ = 45.3 and  $\lambda_{\rm Edd}$ $= 0.13$, broadly similar to RBS 1124. It also exhibits comparable spectral characteristics to RBS 1124, including a soft excess and a narrow unresolved iron emission line, but no obvious detections of wind features in the X-ray spectrum \citep{2010ApJ...725.2444K}. Nevertheless, RBS 1055 showed blue-shifted O[III] lines in the optical spectrum, suggesting outflows in the Narrow Line Region (NLR), possibly driven by winds from the accretion disc \citep{2022A&A...666A.169M}. Further analysis through optical spectroscopy can provide additional insights into the presence or absence of outflows in RBS\,1124, but is beyond the scope of this work.

\section*{Conclusions}

We have performed a detailed spectral analysis of new, broadband (0.3 -- 70 keV) X-ray data of the `bare' AGN RBS\,1124 obtained by \textit{XMM-Newton} and \textit{NuSTAR} in coordination. With the help of the broad coverage and high S/N data from \textit{XMM-Newton} and \textit{NuSTAR}, we are able to place robust constraints on various properties of the source.
    \begin{enumerate}
    \item The spectral analysis reveals a moderately hard power-law continuum with $\Gamma \sim$ 1.8 along with a soft excess below 2 keV, a weak neutral iron line, and curvature at high energies. The high-density relativistic reflection model effectively explains the overall broadband spectral continuum, while the iron line was attributed to a distant reflection.
    \item Both variants of the high-density relativistic reflection models, \textit{xillverCp} and \textit{reflionx}, fit the data well. However they provide different spin constraints: $a^* > 0.835$ for Model 1 and $a^* > 0.236$ for Model 2. This discrepancy is driven by differences in the fine details of the calculations included in each model, which result in different reflection continuum fluxes at low energies ($< 1$ keV). In turn, this results in a different relative contribution to the overall flux of the soft excess from low-energy emission lines, and thus differing degrees of relativistic blurring required to produce a smooth soft excess.
    \item These models require RBS 1124 to have a high disc density, log$(n_{\rm e}$/cm$^{-3}$) $\geq 19.2$. Despite high density, RBS 1124 broadly adheres to the anti-correlation between log(n$_{\rm e}$/cm$^{-3})$ and log($m_{\rm BH}\dot{m}^{2}$) expected for a radiation pressure dominated accretion disc. More than 90\% of the disc accretion power is inferred to be dissipated to the corona. 
    \item Our estimates of a low reflection fraction from the source ($R < 0.5$)  are consistent with the findings of \cite{2010MNRAS.401.1315M} and may suggest the presence of an outflowing corona.
    \item Despite a moderately high mass accretion rate, no detectable wind signatures are observed in the X-ray spectrum of RBS 1124. We speculate that this absence could be attributed to the face-on viewing angle of the source. Further studies of the source through optical spectroscopy could reveal additional insights into its outflows.
     \end{enumerate}

\section*{Acknowledgements}

We thank the referee for their useful suggestions that helped us improve the clarity of the paper. JJ acknowledges support from the Leverhulme Trust, the Isaac Newton Trust and St Edmund’s College, University of Cambridge. CR acknowledges support from Fondecyt Regular grant 1230345 and ANID BASAL project FB210003. PK acknowledges the support provided by NASA through the NASA Hubble Fellowship grant HST-HF2-51534.001-A awarded by the Space Telescope Science Institute, which is operated by the Association of Universities for Research in Astronomy, Incorporated, under NASA contract NAS5-26555
 . This research has made use of data obtained with \textit{NuSTAR}, a project led by Caltech, funded by NASA and managed by the NASA Jet Propulsion Laboratory (JPL), and has utilized the \textsc{nustardas} software package, jointly developed by the Space Science Data Centre (SSDC; Italy) and Caltech (USA). This research has also made use of data obtained with \textit{XMM–Newton}, an ESA science mission with instruments and contributions directly funded by ESA Member States. 

\section*{Data Availability}

All the data utilized in this article are publicly available through ESA's \textit{XMM-Newton} Science Archive (\url{https://www.cosmos.esa.int/web/xmm-newton/xsa}) NASA’s HEASARC archive (\url{https://heasarc.gsfc.nasa.gov/}).



\bibliographystyle{mnras}
\bibliography{references} 



\appendix

\section{Some extra material}
\label{Appendix}

\begin{figure*}
\centering
\includegraphics[scale=0.37]{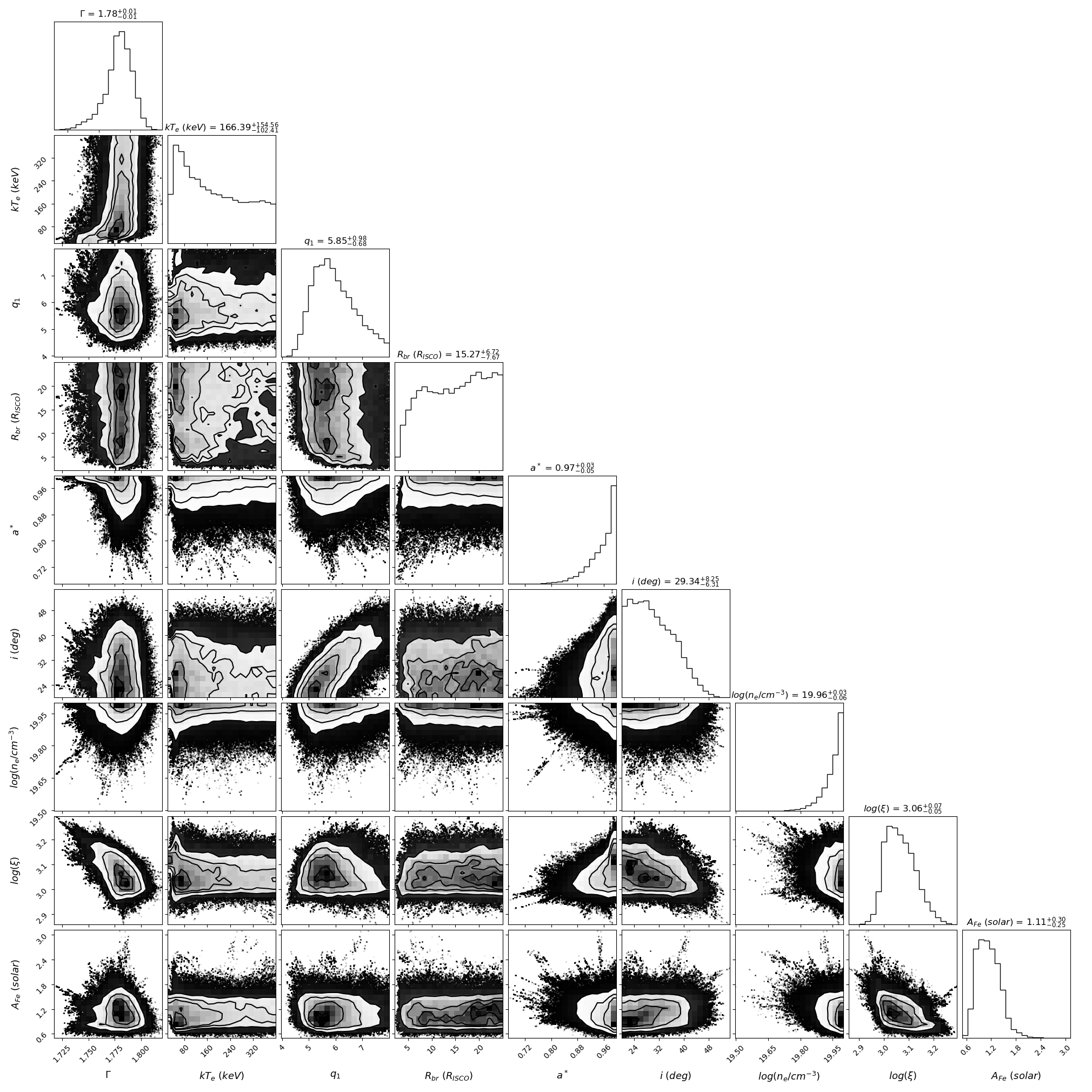}
\caption{The corner plot shows the posterior probability distribution for the parameters from Model 1 obtained from the MCMC analysis using the GW algorithm. The contours in the figure enclose 68\%, 95\%, and 99.7\% of the probabilities.}
\label{fig-A1}
\end{figure*}

\begin{figure*}
\centering
\includegraphics[scale=0.37]{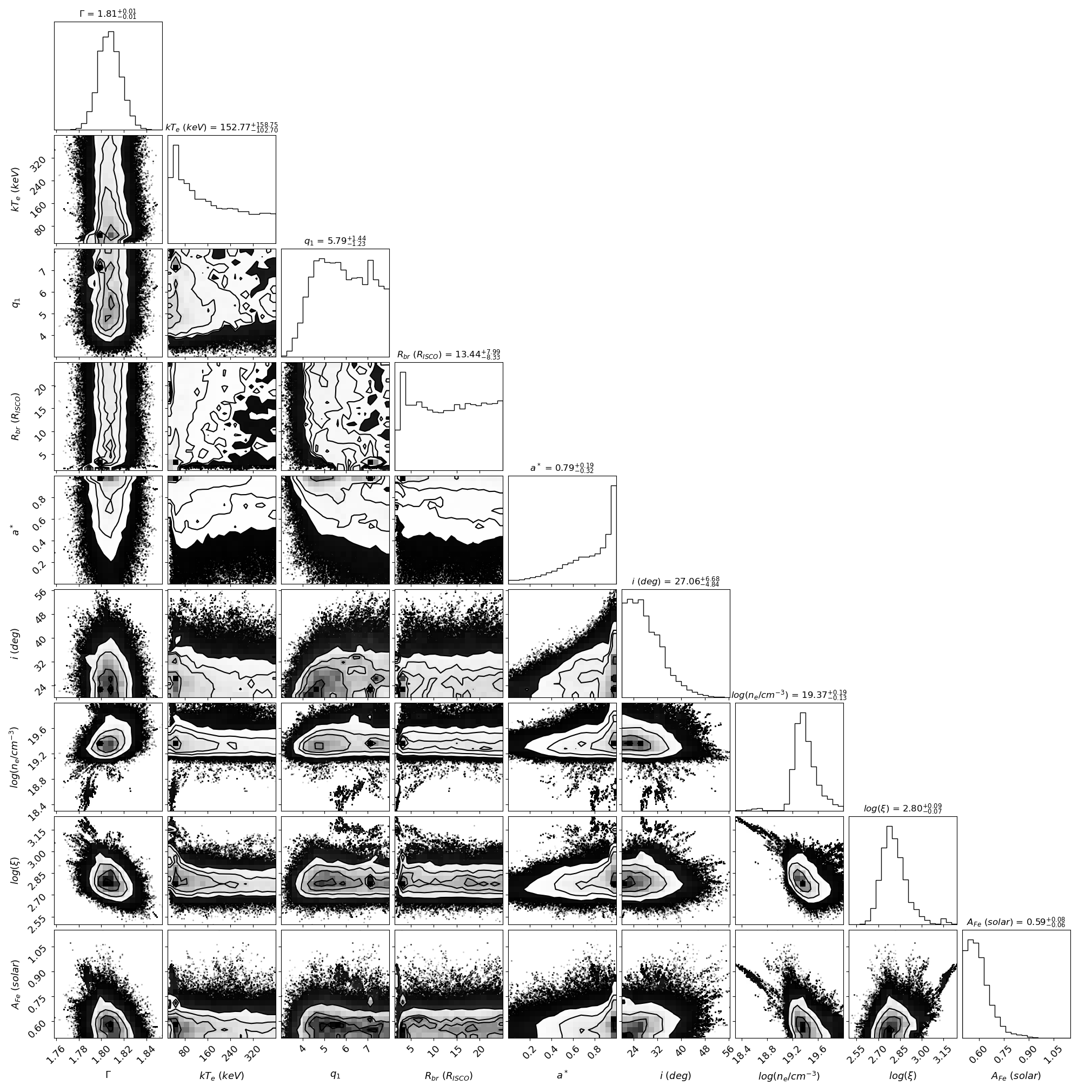}
\caption{The corner plot shows the posterior probability distribution for the parameters from Model 2 obtained from the MCMC analysis. The contours signify the same confidence intervals as quoted in Figure \ref{fig-A1}}
\label{fig-A2}
\end{figure*}

Figures \ref{fig-A1} and \ref{fig-A2} show the MCMC analysis results for the parameter combinations from Model 1 and Model 2, respectively. The specifications of our MCMC sampling approach are outlined in Section \ref{sec-4}. The uncertainty ranges for parameters derived from MCMC are consistent with those estimated by the \textsc{error} command in \textsc{xspec}.

\bsp	
\label{lastpage}
\end{document}